\documentclass[journal]{IEEEtran}
\ifCLASSINFOpdf

\else
\fi
\usepackage{amsmath}
\usepackage{amssymb}
\usepackage{graphicx,amsmath,amssymb,cite}
\usepackage{color}
\usepackage{amsfonts}
\usepackage{amsmath}
\usepackage{graphicx}
\usepackage{amssymb}
\usepackage{stfloats}
\usepackage{subfig}
\usepackage{mathrsfs}
\usepackage{algorithm}
\usepackage{algorithmicx}
\usepackage{algpseudocode}
\usepackage{amsmath}
\usepackage{booktabs}
\usepackage[numbers,sort&compress]{natbib}
\usepackage{comment}
\usepackage{booktabs}
\usepackage[hyphens]{url}

\newcommand{\bm}[1]{\mbox{\boldmath{$#1$}}}

\usepackage{etoolbox}
\apptocmd{\maketitle}{%
  \markboth{The definitive version was published in IEEE Transactions on Wireless Communications, vol. 25, pp. 11932-11948, 2026, doi: 10.1109/TWC.2026.3662428.}{}%
}{}{}

\begin{document}
\title{A Fingerprint Database Generation Method for RIS-Assisted Indoor Positioning}

\author{Xin Cheng, Yu He, Menglu Li, Ruoguang Li, Feng Shu, 
 and Guangjie Han,~\IEEEmembership{Fellow,~IEEE}

\thanks{Xin Cheng, Yu He, Ruoguang Li and Guangjie Han are with the College of Information Science and Engineering, Hohai University, Changzhou 213200, China. (e-mail: xincstar23@163.com).}
\thanks{Menglu Li is with the College of Electrical and Power Engineering, Hohai University, Changzhou 213200, China.}
\thanks{
Feng Shu is with the School of Information and Communication Engineering,
Hainan University, Haikou 570228, China.}}

\maketitle
\begin{abstract}
Reconfigurable intelligent surface (RIS) has emerged as a promising technology to enhance indoor wireless communication and sensing performance. However, the construction of reliable received signal strength (RSS)-based fingerprint databases for RIS-assisted indoor positioning remains an open challenge due to the lack of realistic and spatially consistent channel modeling methods. In this paper, we propose a novel method with open-source code for generating RIS-assisted RSS fingerprint databases. Our method captures the complex RIS-assisted multipath behaviors by extended cluster-based channel modeling and the physical and electromagnetic properties of RIS and transmitter (Tx). And the spatial consistency is incorporated when simulating the fingerprint data collection across neighboring positions. Moreover, an effective sorting algorithm is proposed to solve the online synchronization issue, a closed-form RIS phase configuration strategy is proposed to improve the localization accuracy, and the modeling method of mutual coupling (MC) effect is provided. Extensive simulations are conducted to evaluate the fingerprint database generated by the proposed method. And the positioning performance on the database using different algorithms is analyzed, providing valuable insights for the system design.
\end{abstract}

\IEEEpeerreviewmaketitle

\section{Introduction}
Indoor positioning has become a cornerstone technology in various applications such as indoor navigation, asset tracking, smart homes, and industrial automation \cite{zafari2019survey, nkrow2024nlos}. These applications rely on precise and reliable positioning systems to accurately determine the location of users or devices within complex indoor environments, where traditional global positioning systems (GPS) encounter limitations due to signal obstruction caused by buildings or other structures \cite{singh2024machine}. In response to these challenges, indoor wireless positioning has emerged as a highly effective solution. This approach typically relies on indoor access points (APs) emitting wireless signals such as WiFi and millimeter wave (mmWave) \cite{shastri2022review,gonzalez2024integrated}, which offer adaptability to indoor environments, ease of deployment, and scalability \cite{ cheng2022federated, e2025systematic}.

Indoor wireless positioning does not always perform well, especially in complex environments with significant obstacles such as thick walls, heavy furniture, or other dense materials. Although wireless signals are emitted from an indoor AP, their poor penetration and high susceptibility to blockage make them highly sensitive to environmental conditions \cite{wu2023fingerprint}. In obstructed indoor spaces, wireless signals can experience severe signal attenuation and multipath fading, leading to a significant degradation in positioning accuracy.  To overcome these challenges, the reconfigurable intelligent surface (RIS) has emerged as a promising solution. RIS can dynamically reconfigure the radio environment by reflecting, refracting, or diffracting signals, effectively mitigating the effects of obstacles and improving signal propagation quality \cite{wu2019towards,cheng2021joint,shu2024precoding}, thereby significantly  enhancing positioning accuracy  \cite{emenonye2023fundamentals}.

To achieve RIS-assisted indoor position sensing, two main approaches are typically employed. One is the geometric-based wireless positioning, which relies on channel parameter estimation, and the other is the fingerprint-based wireless localization (WFL), which depends on data collection. The geometric-based method usually consists of two steps \cite{wang2021joint, pan2023ris,zhang2023approximate}. First, the RIS-assisted channel parameters, such as angles of arrival (AoA) or time of flight (ToF), are estimated. Then the user's position is calculated based on geometric relationships. Although effective, this method is computationally intensive and may face accuracy limitations due to its reliance on channel model assumptions. On the other hand, the fingerprint-based method consists of offline phase and online phase \cite{nguyen2021wireless,zhang2022multiple,wu2023fingerprint,wang2023intelligent,sardellitti2024ris,javed2024fingerprinting,hou2025indoor}. In the offline phase, wireless fingerprint like received signal strength (RSS) data under different RIS configurations are collected within the indoor space, and the fingerprint database is constructed. Then, in the online phase, the real-time position of user is predicted using the fingerprint database or trained neural network model. In this approach, different RIS configurations generate various radio map of the indoor space, resulting in large spatiotemporal freedom for wireless positioning \cite{wang2023intelligent,sardellitti2024ris}. Compared to the first one, the WFL is less sensitive to the complexities of the channel and environmental assumptions, offering a higher accuracy and robustness.

Existing works on RIS-assisted WFL mainly researched high-accuracy localization algorithms and effective RIS configurations. Among them, K-nearest neighbors (KNN)  \cite{wu2023fingerprint,sardellitti2024ris,hou2025indoor}, deep neural network (DNN) \cite{zhang2022multiple} and residual convolution network regression learning-based algorithm \cite{wu2023fingerprint} were studied and demonstrated on RIS-assisted scenarios.
Moreover, the RIS configurations, i.e., the phase shifts of RIS are optimized by supervised learning-based feature selection  \cite{nguyen2021wireless},  projected gradient descent \cite{zhang2022multiple},  deep reinforcement learning \cite{wang2023intelligent} and  multilayer graph representations \cite{sardellitti2024ris}. Simulation results showed that these approaches generate effective RIS-assisted radio maps, thus improving the positioning accuracy.

The verification and evaluation of existing works on RIS-assisted WFL  were conducted on the synthetic RSS fingerprint database \cite{nguyen2021wireless,zhang2022multiple,wu2023fingerprint,wang2023intelligent,sardellitti2024ris,javed2024fingerprinting,hou2025indoor}, as constructing real-world fingerprint database requires time-consuming manual measurement campaigns and high-cost hardware. However, the RIS-assisted indoor wireless propagation used for generating the fingerprint database was clearly  simplified in these works. In \cite{nguyen2021wireless,wang2023intelligent,hou2025indoor}, only the virtual line-of-sight (VLoS) path from RIS is considered when simulating the wireless propagation. In \cite{zhang2022multiple,sardellitti2024ris,javed2024fingerprinting}, only VLoS path and line-of-sight (LoS) path are considered. Multipath components caused by indoor obstacles are simplified to a Gaussian noise. Moreover, due to the complex indoor propagation environment, LoS and VLoS paths may be obstructed at many positions in the indoor space. Authors in \cite{wu2023fingerprint} used a software named Wireless Insite to simulate a relatively real RIS-assisted indoor wireless propagation. Authors in \cite{javed2024fingerprinting} also proposed a database generation method called FWS-E2E, which employs a commercial full-wave electromagnetic (EM) simulator such as CST or HFSS. However, these software are closed-source and expensive, and the construction  flexibility based on them is limited.

Motivated by the aforementioned gaps, this paper proposes a novel open-source RIS-assisted RSS fingerprint database construction method. To capture the complex reflections, diffractions, and scattering behaviors typical in indoor environments, cluster-based channel modeling \cite{3GPP_TR_38_901,poddar2023tutorial} is used and extended to the RIS-assisted scenario, in which the multipath components in the indoor environment are grouped into clusters. Moreover, to ensure smooth RSS variations across nearby locations and enhance the reliability of the fingerprint database, the spatial consistency  property is considered, which has not been addressed in existing work on RIS channel modeling \cite{basar2021indoor,xiong2021statistical,lian2024physics,yuan2024ris}.
The main contributions and novelties of this paper are summarized as follows:
\begin{enumerate}
\item We propose a general and flexible channel modeling method for characterizing  complex indoor RIS-assisted wireless propagation in the presence of multiple clusters and scatterers. This method accounts for the physical and electromagnetic properties of RIS and antennas, such as the number of units, size, position, orientation, and directional/omnidirectional gain. The proposed channel model is versatile and can be applied to describe a wide range of RIS-assisted communication and sensing scenarios, beyond RIS-assisted WFL.
\item To ensure more realistic and correlated channel modeling across different indoor locations, we introduce a spatial consistency procedure for RIS-assisted indoor wireless propagation, which contributes to an overall indoor channel model. Building on this, we propose a RSS fingerprint generation method accompanied by open-source code. Accurate and tailored RIS-assisted RSS fingerprint databases can be generated by adjusting the parameters to suit specific scenarios.
\item Practical issues of the RIS-assisted WFL are addressed. To tackle the issue of online measurement synchronization, a simple and efficient sorting algorithm is proposed. To enhance the diversity of radio maps in WFL, a closed-form empirical strategy for RIS phase configurations is proposed.  Furthermore, the often-overlooked mutual coupling (MC) effect is considered and the general modeling method is provided.
\item The generated fingerprint databases and the corresponding positioning performance are analyzed. Extensive simulations validate the effectiveness of integrating RIS in assisting WFL. Furthermore, the simulation results highlight key factors influencing positioning accuracy, including the physical and electromagnetic properties of RIS, the phase configurations of RIS, and measurement configurations, offering valuable insights for real-world design considerations.
\end{enumerate}

The remainder of this article is organized as follows. In Section II, the framework of RIS-assisted WFL is illustrated, and then the problem of generating fingerprint database is introduced. In Section III, a general cluster-based RIS-assisted indoor channel modeling method is proposed. Then, a spatial consistency procedure is developed for the overall indoor channel modeling in Section IV. Section V illustrates the proposed fingerprint database generation method. Section VI discusses and solves three practical issues of the RIS-assisted WFL.
Section VII shows the simulation setup and results. Finally, Section VIII concludes the paper.

\emph{Notations:} Boldface lower case and boldface upper case letters denote vectors and matrices, respectively. Sign $(\cdot)^{T}$ denotes the transpose operation. Sign $|| \cdot ||$ denotes the Frobenius norm of a matrix or the L2 norm of a vector.

\section{RIS-assisted WFL Framework}\label{smpssection}
\begin{figure}
  \centering
  \includegraphics[width=0.5\textwidth]{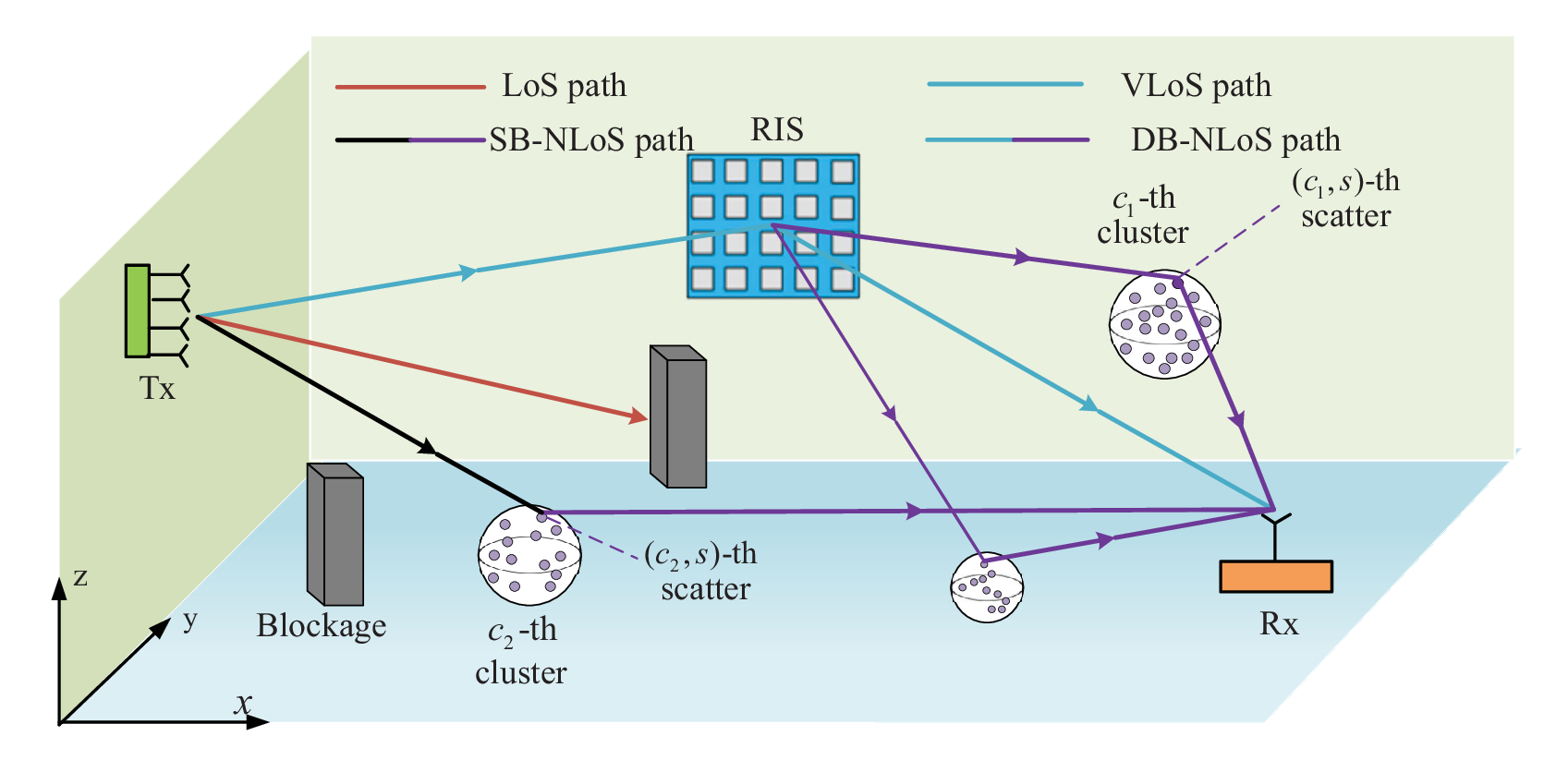}\\
  \caption{RIS-assisted indoor wireless propagation.}\label{riswtfig}
\end{figure}

Let us consider the RIS-assisted WFL in an indoor wireless propagation environment, as shown in Fig.~\ref{riswtfig}, where a multi-antenna AP serves as the transmitter (Tx) and a RIS is placed to modulate the wireless propagation through phase configurations on different reflective units \cite{nguyen2021wireless}. Affected by indoor objects and modulated by the RIS, the wireless signal transmitted by the Tx can form an indoor radio map, which can be measured by a device equipped with a RSS sensor, acting as the receiver (Rx).

The RIS-assisted WFL operates in the offline phase and the online phase. During the offline phase, fingerprint databases are generated through a site survey in the area of interest (AoI). Specifically, the Rx measures RSS values at various indoor surveying positions \cite{alitaleshi2023ea}. At each surveying position, multiple measurements are taken while different RIS configurations are set. Consequently, multiple RSS values along with the recorded surveying coordinate are collected to form a fingerprint data pair for each surveying position. By collecting all fingerprint data pairs, a fingerprint database is then generated. This database is then transmitted to a network operator. On the network operator's side, the fingerprint database is either stored or utilized as a training dataset to generate a neural network-based localization model.
In the online phase, any device, besides the one doing the site surveying, can require real-time positioning by simply sending its measured RSS values under different RIS configurations to the network operator.  The network operator then estimates the device's position and sends the result back to it.

Following the above RIS-assisted WFL framework, a fingerprint database generation method will be proposed in next sections. For convenience, some commonly-used notations are listed in Table \ref{tabecnotations}. The symbols in Table \ref{tabecnotations} are also defined formally where they first appear in the paper.

\begin{table}
\centering
\captionsetup{justification=centering, labelsep=newline,textfont=sc}
\caption{Commonly Used Notations}
\label{tabecnotations}
\begin{tabular}{p{0.1\textwidth}|p{0.3\textwidth}}
\hline
\multicolumn{1}{c|}{\textbf{Symbol}} & \multicolumn{1}{c}{\textbf{Definition}} \\
\hline
$\lambda$ & Wavelength \\
\hline
$M_{T}$,  $I$ & Numbers of Tx antennas and reflective units\\
\hline
$M_{I}$, $N_{I}$ & Numbers of columns and rows of RIS\\
\hline
$N$ & Number of measurments \\
\hline
$\mathbb{C}_{1}, \mathbb{C}_{2}$ & Index sets of clusters in SB-NLoS link and DB-NLoS link \\
\hline
$C_{1}, C_{2}$ & Numbers of clusters in SB-NLoS link and DB-NLoS link \\
\hline
$S_{c_{1}},S_{c_{2}}$ & Numbers of intra-cluster scatterers in the $c_{1}$-th cluster and the $c_{2}$-th cluster \\
\hline
$\bm{\Omega}$ & Phase shift matrix of RIS \\
\hline
$\theta_{i}$ & Phase shift of the $i$-th reflective unit \\
\hline
$A$  &  Reflection magnitude of each reflective unit \\
\hline
$\zeta_{\mathrm{LoS}}, \zeta_{\mathrm{VLoS}}$ & Availability indicators of LoS path and VLoS path \\
\hline
$L^{\mathrm{LoS}}$, $L^{\mathrm{VLoS}}$ &  Link attenuations of LoS path and VLoS path\\
\hline
$L^{\mathrm{SB-NLoS}}$, $L^{\mathrm{DB-NLoS}}$ & Link attenuations of SB-NLoS path and  DB-NLoS path\\
\hline
$n_{\mathrm{LoS}}$, $n_{\mathrm{NLoS}}$  & Path loss coefficients of LoS path and NLoS path \\
\hline
$\chi_{\mathrm{LoS}}$, $\chi_{\mathrm{NLoS}}$&  SF terms of LoS path and NLoS path \\
\hline
$G_{T}$, $G_{R}$, $G_{I}$ & Radiation patterns of Tx antenna, Rx antenna and reflective unit\\
\hline
$\Delta d^{T}, \Delta d^{I}_{x}, \Delta d^{I}_{y}$ & Antenna spacing of Tx, length and
width of a reflective unit \\
\hline
\end{tabular}
\end{table}

\section{RIS-Assisted Indoor Channel Modeling}
In this section, we propose a general RIS-assisted indoor channel modeling method as a cornerstone to generate a fingerprint database, as described in Section \ref{secfdgf}. First, we outline the basic settings for RIS-assisted indoor wireless propagation in the presence of multiple clusters and scatterers. Next, the channel impulse response (CIR) is modeled by integrating four types of propagation paths. The calculation of the channel parameters in the CIR is outlined next. Finally, a statistical method for generating clusters and scatterers is extended to the RIS-assisted scenario.

\subsection{Basic Settings}
In the RIS-assisted indoor wireless propagation environment, as shown in Fig.~\ref{riswtfig}, the Tx is equipped with a uniform linear array (ULA) and the Rx is equipped with a single antenna. A rectangular RIS is also considered, consisting of \( I \) reflective units. The number of Tx antennas is denoted as \( M_T \), and the antenna spacing of Tx is denoted as \( \Delta d^T \).  Let \( M_I \) and \( N_I \) denote the number of columns (counted from left to right) and rows (counted from bottom to top) of the regularly arranged reflective units in the RIS array, respectively. The length and width of each reflective unit are denoted as \( \Delta d^I_x \) and \( \Delta d^I_y \), respectively. The size of each reflective unit is typically on the sub-wavelength scale \cite{xiong2021statistical}.

\begin{figure}
  \centering
  \includegraphics[width=0.5\textwidth]{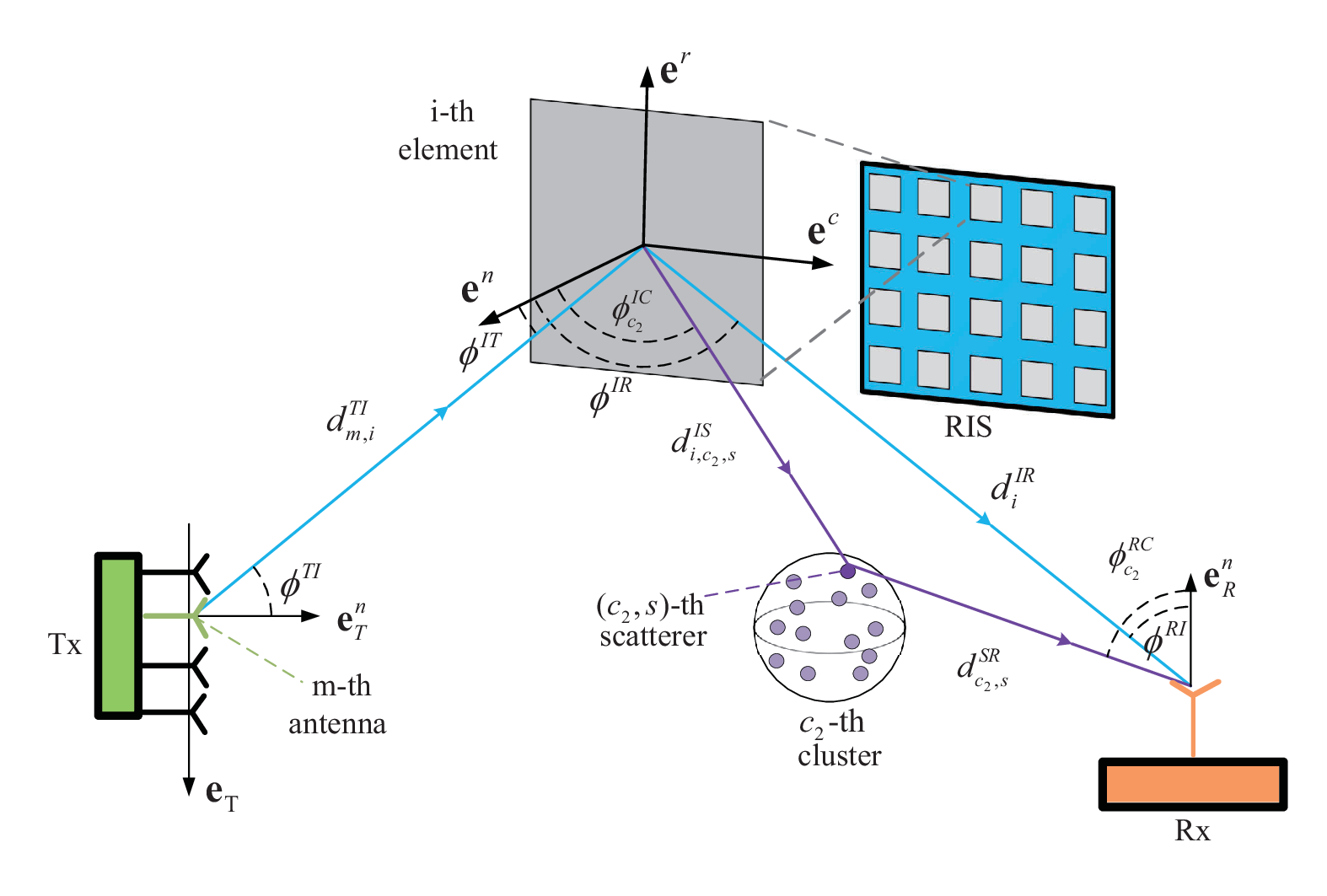}\\
  \caption{Illustration of some orientations, angles and distances.}\label{georelafig}
\end{figure}

As shown in Fig.~\ref{riswtfig}, a three-dimensional (3-D) Cartesian coordinate system is established, with the origin placed within the indoor environment. The Tx is fixed at position \( \bm{\xi}_T \). Similarly, the RIS is fixed at position \( \bm{\xi}_I \). The Rx is denoted as \( \bm{\xi}_R \). Note that the Rx position changes when collecting fingerprint data in the offline phase of RIS-assisted WFL. In addition to the positions of Tx, RIS and Rx, their orientations also play an important role in determining the wireless propagation characteristics \cite{zeng2020reconfigurable}, which are often overlooked in existing studies. In this paper, we define the orientations using directional vectors and normal vectors within the Cartesian coordinate system, as shown in Fig.~\ref{georelafig}. For Tx, the directional vector  is the unit direction vector of ULA, while the normal vector  is the unit pointing vector of antenna. Let \( \mathbf{e}_T \) and \( \mathbf{e}_T^n \) denote the directional vector and normal vector of Tx, respectively. Let \( \mathbf{e}_R^n \) denote the normal vector of Rx.
For the RIS, the directional vectors consist of two unit vectors, one is the unit row vector \( \mathbf{e}^r \), and the other is the unit column vector \( \mathbf{e}^c \). The normal vector of the RIS is the unit vector perpendicular to the RIS plane, denoted as \( \mathbf{e}^n \). We have the  following orthogonal relations: $\mathbf{e}^r \perp \mathbf{e}^c \perp \mathbf{e}^n$.

In the indoor environment, there are many objects, such as tables and chairs  scattering or reflecting the wireless signal. To describe the impact of indoor objects on wireless propagation, cluster-based models have been proposed by 3rd generation partnership project (3GPP)  \cite{3GPP_TR_38_901} and New York University \cite{poddar2023tutorial}, where each cluster consisting of many scatterers accounts for a distinguishable path with resolvable delay.  Normally, multiple paths involving clusters degrade the signal quality. To avoid this influence artificially, the Tx and RIS are placed at hand-picked positions so that there are no clusters in the Tx-RIS link. This is possible since the Tx and RIS are controllable. Since the Rx should traverse indoor positions during the site survey and the online servicing devices of RIS-asssited WFL can appear at any position, the clusters in Tx-Rx link and RIS-Rx link are inevitable, thus considered in this paper as shown in Fig.~\ref{riswtfig}.

Let $\mathbb{C}_{1}$ denote the cluster set between the Tx and Rx and $\mathbb{C}_{2}$ denote the cluster set between the RIS and Rx.  The number of clusters in $\mathbb{C}_{1}$  is denoted as $C_{1}$, while the number of clusters in $\mathbb{C}_{2}$ is denoted as $C_{2}$. Assume there are $S_{c_{1}}$ scatterers in the $c_{1}$-th cluster in $\mathbb{C}_{1}$, and $S_{c_{2}}$ scatterers in the $c_{2}$-th cluster in $\mathbb{C}_{2}$. Let $\bm{\xi}_{c_{1}}$ denote the position of the $c_{1}$-th cluster in $\mathbb{C}_{1}$, while $\bm{\xi}_{c_{2}}$ denote the position of the $c_{2}$-th cluster in $\mathbb{C}_{2}$. The position of the $(c_{1},s)$-th scatterer, i.e., the $s$-th scatterer in the $c_{1}$-th cluster, is denoted as $\bm{\xi}_{c_{1},s}$. Similarly, the position of the $(c_{2},s)$-th scatterer is denoted as $\bm{\xi}_{c_{2},s}$.

\subsection{Channel Impulse Response}\label{cirsubsection}
As illustrated in Fig.~\ref{riswtfig}, the wireless propagation paths between the Tx and Rx are divided into four paths. The first is the direct link between the Tx and Rx,  i.e., Tx$\rightarrow$Rx, which is referred  to as the LoS path. The second is the link involving the Tx, RIS, and Rx,  i.e., Tx$\rightarrow$RIS$\rightarrow$Rx, referred to as the VLoS path. The third is the link involving the Tx, clusters, and Rx, i.e., Tx$\rightarrow\mathbb{C}_1\rightarrow$Rx, known as the single-bounced non-line-of-sight (SB-NLoS) path \cite{xiong2021statistical}. The fourth is the link involving the Tx, RIS, clusters, and Rx,  i.e., Tx$\rightarrow$RIS$\rightarrow\mathbb{C}_2\rightarrow$Rx, known as the double-bounced non-line-of-sight (DB-NLoS) path \cite{xiong2021statistical}. Note that paths affected by multiple clusters are excluded from consideration as in \cite{xiong2021statistical,hemadeh2017millimeter}. Because higher-order reflections and diffractions experience significant power attenuation, their contribution to the total received signal becomes negligible, compared to the aforementioned paths.

Due to the multiple-antennas property of Tx, the channel model can be characterized by
a complex matrix of size $1\times M_{T}$, that is, $\mathbf{H}(\tau)=\begin{bmatrix} h_{1}(\tau) & h_{2}(\tau)  & \cdots &  h_{M_{T}}(\tau) \end{bmatrix}$, where $h_{m}$ denotes the CIR between the m-th Tx antenna and the Rx antenna with $m=1,2, \cdots, M_{T}$, and $\tau$ stands for the path delay from the Tx to the Rx. Furthermore, $h_{m}(\tau)$ is the summation of the LoS, VLoS, SB-NLoS, and DB-NLoS components, given by
\begin{align}\label{totallchannel}
h_{m}(\tau)=&\zeta_{\mathrm{LoS}}h^{\mathrm{LoS}}_{m}(\tau)+\zeta_{\mathrm{VLoS}}h^{\mathrm{VLoS}}_{m}(\tau)\\ \nonumber
&+h^{\mathrm{SB-NLoS}}_{m}(\tau)+h^{\mathrm{DB-NLoS}}_{m}(\tau),
\end{align}
where $h^{\mathrm{LoS}}_{m}$, $h^{\mathrm{VLoS}}_{m}$, $h^{\mathrm{SB-NLoS}}_{m}$, and $h^{\mathrm{DB-NLoS}}_{m}$ stand for the CIRs between the $m$-th Tx antenna and the Rx antenna belonging to the LoS path, the VLoS path, the SB-NLoS path and the DB-NLoS path, respectively, and $\zeta_{\mathrm{LoS}}$ represents the availability indicator of the LoS path while $\zeta_{\mathrm{VLoS}}$ represents the availability indicator of the VLoS path. When the direct path between Tx and Rx  is not blocked,  $\zeta_{\mathrm{LoS}}=1$, otherwise, $\zeta_{\mathrm{LoS}}=0$. Since there is a direct link between Tx and RIS, when the direct path between RIS and Rx is not blocked, $\zeta_{\mathrm{VLoS}}=1$, otherwise,  $\zeta_{\mathrm{VLoS}}=0$.

According to the wireless propagation principles \cite{molisch2012wireless,basar2021indoor},  the LoS component of the CIR between the $m$-th Tx antenna and the Rx antenna is given by \cite{3GPP_TR_38_901,poddar2023tutorial}
\begin{equation}\label{channellos}
h^{\mathrm{LoS}}_{m}(\tau) = \sqrt{L^{\mathrm{LoS}} G_T(\phi^{TR}) G_R(\phi^{RT})} e^{-j2\pi \frac{d^{TR}_m}{\lambda}} \delta(\tau - \tau^{\mathrm{LoS}}),
\end{equation}
where \( L^{\mathrm{LoS}} \) stands for the link attenuation caused by the LoS path loss, \( \phi^{TR} \) denotes the elevation angle from Tx to Rx while \( \phi^{RT} \) denotes the elevation angle from Rx to Tx, \( G_T \) is the radiation pattern of the Tx antenna, \( G_R \) is the radiation pattern of the Rx antenna, \( d^{TR}_m \) is the distance between the $m$-th Tx antenna and the Rx, \( \lambda \) is the wavelength of the signal, and \( \tau^{\mathrm{LoS}} \) is the time delay of this path, and \(\tau^{\mathrm{LoS}} = \frac{d^{TR}}{c}\), where \( c \) is the speed of light and \( d^{TR} \) is the distance between Tx and  Rx.

In the VLoS path, the wireless signal emitted from the Tx to the Rx is reflected by the reflective units of the RIS. Each controllable reflective unit applies an independent phase shift to the incident EM waves, and EM waves reflected by different reflective units are integrated at the Rx broadside. Considering the physics and electromagnetic nature of the RIS \cite{tang2022path,xiong2021statistical,basar2021indoor}, the impact of RIS can be divided into three parts: one is the reflective path loss caused by the RIS; another is the radiation pattern of reflective units; the other is the phase modulation caused by reflective units. Based on the above, the VLoS component of the CIR between the $m$-th Tx antenna and the Rx antenna is expressed as \cite{tang2022path,xiong2021statistical,basar2021indoor}
\begin{align}\label{channelvlos}
h_m^{\mathrm{VLoS}}(\tau)=&\sqrt{L^{\mathrm{VLoS}} G_T(\phi^{TI}) G_R(\phi^{RI}) G_I(\phi^{IT})
 G_I(\phi^{IR})} \\\nonumber
 & \mathbf{a}_{m}^{TI}\bm{\Omega}\mathbf{a}^{IR}\delta \left( \tau-\tau^{\mathrm{VLoS}}\right)
\end{align}
where \( L^{\mathrm{VLoS}} \) stands for the link attenuation caused by the VLoS path loss, \( \phi^{IT} \) denotes the elevation angle from RIS to Tx while \( \phi^{IR} \) denotes the elevation angle from RIS to Rx, \( G_I\) is the radiation pattern of the reflective unit, $\mathbf{a}_{m}^{TI}$ is the steering vector from the $m$-th Tx antenna to RIS,  $\mathbf{a}^{IR}$ is the steering vector from  RIS to Rx,
$\bm{\Omega}$ is the phase shift matrix of RIS, and \(\tau^{\mathrm{VLoS}}\) represents the time delay of the VLoS path given by  \(\tau^{\mathrm{VLoS}} = \frac{d^{TI} + d^{IR}}{c}\), where \( d^{TI} \) is the distance from Tx to RIS, and \( d^{IR} \) is the distance from RIS to Rx. Note that in Eq. (\ref{channelvlos}), the link attenuations or radiation patterns among different reflective units are assumed to be uniform, because the length/width of the reflective unit is much smaller than the distance between the RIS and Tx/Rx.  $\bm{\Omega}$ is a diagonal matrix whose $i$-th diagonal element represents the phase shift of the $i$-th reflective unit, denoted as $\theta_{i}$. Moreover, the steering vectors are given by
\begin{subequations}
\begin{align}
\mathbf{a}_{m}^{TI}=\begin{bmatrix} e^{-j2\pi \frac{d^{TI}_{m,1}}{\lambda} }  &  e^{-j2\pi \frac{d^{TI}_{m,2}}{\lambda} }  & \cdots &  e^{-j2\pi \frac{d^{TI}_{m,I}}{\lambda} } \end{bmatrix},
\end{align}
\begin{align}
\mathbf{a}^{IR}=\begin{bmatrix} e^{-j2\pi \frac{d^{IR}_1}{\lambda} }  &  e^{-j2\pi \frac{d^{IR}_2}{\lambda} }  & \cdots &  e^{-j2\pi \frac{d^{IR}_I}{\lambda} } \end{bmatrix},
\end{align}
\end{subequations}
\( d^{TI}_{m,i} \) represents the distance from the $m$-th Tx antenna to the $i$-th reflective unit, \( d^{IR}_i \) represents the distance from the $i$-th reflective unit to the Rx.

The SB-NLoS path characterizes the multipath propagation from the Tx to the Rx via single-bounced
reflections through clusters in $\mathbb{C}_{1}$. Note that the SB-NLoS path is independent of the RIS. The SB-NLoS component of the CIR between the $m$-th Tx antenna and the Rx is expressed as \footnote{Unlike channel models that use the K-factor \cite{3GPP_TR_38_901}, the channel model in this paper is expressed by integrating all sub-paths with absolute path strength, and therefore no normalization is applied. }\cite{3GPP_TR_38_901,poddar2023tutorial}
\begin{align}\label{channelsbnlos}
&h_m^{\mathrm{SB-NLoS}}(\tau)= \\\nonumber
& \sum_{c_{1}=1}^{C_1} \sum_{s=1}^{S_{c_{1}}} \beta_{c_{1},s}
\sqrt{L_{c_{1}}^{\mathrm{SB-NLoS}} G_T(\phi^{TC}_{c_{1}}) G_R(\phi^{RC}_{c_{1}})} \\\nonumber
&e^{-j2\pi \frac{d^{TS}_{m,c_{1},s} + d^{SR}_{c_{1},s}}{\lambda}} \delta \left( \tau - \tau_{c_{1},s}^{\mathrm{SB-NLoS}} \right),
\end{align}
where \( L_{c_{1}}^{\mathrm{SB-NLoS}} \) stands for the link attenuation caused by this non-line-of-sight  (NLoS) path involving the $c_{1}$-th cluster in $\mathbb{C}_{1}$, $\beta_{c_{1},s}$  stands for the complex gain of  the $(c_{1},s)$-th scatterer in $\mathbb{C}_{1}$, \( \phi^{TC}_{c_{1}} \) denotes the elevation angle from Tx to the $c_{1}$-th cluster, while \(\phi^{RC}_{c_{1}} \) denotes the elevation angle from Rx to the $c_{1}$-th cluster, \(d^{TS}_{m,c_{1},s}\) and \( d^{SR}_{c_{1},s}\) represent  the distance from the $m$-th Tx antenna to the $(c_{1},s)$-th scatterer and the distance from the $(c_{1},s)$-th scatterer to the Rx, respectively, and $\tau_{c_{1}}^{\mathrm{SB-NLoS}}$ is the time delay of this path with $\tau_{c_{1},s}^{\mathrm{SB-NLoS}} = \frac{d^{TS}_{c_{1},s} + d^{SR}_{c_{1},s}}{c}$, where \( d^{TS}_{c_{1},s}  \) and \( d^{SR}_{c_{1},s} \) represent the distance from Tx to the $(c_{1},s)$-th scatterer and the distance from the  $(c_{1},s)$-th scatterer to Rx, respectively. $\beta_{c_{1},s}=a_{c_{1},s}e^{j\theta_{c_{1},s}}$, where $a_{c_{1},s}\in \mathcal{U}(0,1]$, which is the random loss factor of the $(c_{1},s)$-th scatterer, and  $\theta_{c_{1},s}\in\mathcal{U}[0,2\pi)$, which is the random phase of the $(c_{1},s)$-th scatterer.

The DB-NLoS path characterizes the multipath propagation from the Tx to the Rx via double-bounced
reflections, where the RIS results in the first bounce and the clusters in $\mathbb{C}_{2}$ account for the second bounce. The DB-NLoS component of the CIR between the $m$-th Tx antenna and the Rx is expressed as \cite{xiong2021statistical,basar2021indoor}
\begin{align}\label{channeldbnlos}
&h_m^{\mathrm{DB-NLoS}}(\tau) =\sum_{c_{2}=1}^{C_2} \sum_{s=1}^{S_{c_{2}}} \beta_{c_{2},s} \\\nonumber
&\sqrt{L_{c_{2}}^{\mathrm{DB-NLoS}} G_T(\phi^{TI}) G_R(\phi^{RC}_{c_{2}}) G_I(\phi^{IT}) G_I(\phi^{IC}_{c_{2}})}\\\nonumber
&\mathbf{a}_{m}^{TI}\bm{\Omega}\mathbf{a}^{ICR}_{c_{2},s} \delta \left( \tau - \tau_{c_{2},s}^{\mathrm{DB-NLoS}} \right)
\end{align}
where \( L_{c_{2}}^{\mathrm{DB-NLoS}}\) represents the link attenuations caused by this NLoS path involving the RIS and the $c_{2}$-th cluster in $\mathbb{C}_{2}$, $\beta_{c_{2},s}$ stands for the complex gain of the $(c_{2},s)$-th scatterer in $\mathbb{C}_{2}$, \( \phi^{RC}_{c_{2}}\) denotes the elevation angle from  Rx to the $c_{2}$-th cluster, while \(\phi^{IC}_{c_{2}}\) denotes the elevation angle from  RIS to the $c_{2}$-th cluster, $\mathbf{a}^{ICR}_{c_{2},s}$ is the joint steering vector from  RIS to the $(c_{2},s)$-th scatterer and from the $(c_{2},s)$-th scatterer to the Rx, and $\tau_{c_{2}}^{\mathrm{DB-NLoS}}$ is the time delay of this path with $\tau_{c_{2},s}^{\mathrm{DB-NLoS}} = \frac{d^{TI} + d^{IS}_{c_{2},s} + d^{SR}_{c_{2},s}}{c}$, where \( d^{IS}_{c_{2},s}\)  represents the distance from the RIS to the $(c_{2},s)$-th scatterer and \( d^{SR}_{c_{2},s}\) represents the distances from the $(c_{2},s)$-th scatterer to the Rx. $\beta_{c_{2},s}=a_{c_{2},s}e^{j\theta_{c_{2},s}}$, where $a_{c_{2},s}\in \mathcal{U}(0,1]$, which is the random loss factor of the $(c_{2},s)$-th scatterer, and  $\theta_{c_{2},s}\in\mathcal{U}[0,2\pi)$, which is the random phase of the $(c_{2},s)$-th scatterer. Moreover, the  steering vector is given by
\begin{align}
&\mathbf{a}^{ICR}_{c_{2},s}=\\\nonumber
&\begin{bmatrix}  \!  e^{-j2\pi \frac{d^{IS}_{1,c_{2},s}+d^{SR}_{c_{2},s}}{\lambda} }  \!  & \! e^{-j2\pi \frac{d^{IS}_{2,c_{2},s}+d^{SR}_{c_{2},s}}{\lambda}}  \!  &  \!  \cdots  \!  &  \!  e^{-j2\pi \frac{d^{IS}_{I,c_{2},s}+d^{SR}_{c_{2},s}}{\lambda}}  \!  \end{bmatrix},
\end{align}
where \( d^{IS}_{i,c_{2},s} \)  is the distance from the $i$-th reflective unit to the $(c_{2},s)$-th scatterer, \( d^{SR}_{c_{2},s} \)  is the distance from the $(c_{2},s)$-th scatterer to the Rx. Some angles and distances in the DB-NLoS path are illustrated in Fig.~\ref{georelafig}.

Note that in the above CIR expressions, approximations are made regarding distances and angles. When calculating link attenuations and delays, the central distances from the Tx, Rx and RIS are used in place of the distances between different antennas and reflective units. Similarly, when calculating the radiation pattern, the central angles from the Tx, Rx and RIS are used instead of the angles between different antennas and reflective units. These approximations significantly reduce the complexity of the model. Generally, the deviations of distance approximations are minimal, on the order of the wavelength, thus these deviations can be ignored when calculating link attenuations and delays. Likewise, the deviations of angle approximations are small, and the radiation pattern is not sensitive to these deviations, thus these deviations can be also ignored.

\subsection{Calculation of Channel Model Parameters}
In this subsection, the channel parameters in $h^{\mathrm{LoS}}_{m}$, $h_m^{\mathrm{VLoS}}$, $h_m^{\mathrm{SB-NLoS}}$ and $h_m^{\mathrm{DB-NLoS}}$ are calculated. According to Eq. (\ref{channellos}), Eq. (\ref{channelvlos}), Eq. (\ref{channelsbnlos}) and Eq. (\ref{channeldbnlos}), the undefined  channel model parameters are divided into three types: one  is the geometric relationship including distances and angles; another is the link attenuation; the other is the radiation pattern.

Since the 3-D coordinate of Tx, Rx, RIS and clusters are given, the distances among Tx, Rx, RIS and clusters are calculated directly. We have $d^{TR}=\|\bm{\xi}_{R}-\bm{\xi}_{T}\|$, $d^{TI}=\|\bm{\xi}_{I}-\bm{\xi}_{T}\|$, $d^{IR}=\|\bm{\xi}_{R}-\bm{\xi}_{I}\|$,  $d^{IC}_{c_{2}}=\|\bm{\xi}_{c_{2}}-\bm{\xi}_{I}\|$, $d^{CR}_{c_{2}}=\|\bm{\xi}_{R}-\bm{\xi}_{c_{2}}\|$,$d^{TC}_{c_{1}}=\|\bm{\xi}_{T}-\bm{\xi}_{c_{1}}\|$,
$d^{CR}_{c_{1}}=\|\bm{\xi}_{R}-\bm{\xi}_{c_{1}}\|$,
$d^{TS}_{c_{1},s}=\|\bm{\xi}_{T}-\bm{\xi}_{c_{1},s}\|$,$d^{IS}_{c_{2},s}=\|\bm{\xi}_{I}-\bm{\xi}_{c_{2},s}\|$,
$d^{SR}_{c_{1},s}=\|\bm{\xi}_{R}-\bm{\xi}_{c_{1},s}\|$,
$d^{SR}_{c_{2},s}=\|\bm{\xi}_{R}-\bm{\xi}_{c_{2},s}\|$.
To calculate more meticulous distances involving the specific antenna in Tx or the specific reflective unit in RIS, the distance vector within the Tx antenna array and the distance vector within reflective units are defined. Specifically, the distance vector from the center of Tx antenna array to the m-th Tx antenna, denoted as $\mathbf{A}^{T}_{m}$, is expressed as
\begin{align}
\mathbf{A}^{T}_{m}=\left(m-\frac{M_{T}+1}{2}\right)\Delta d_{T}\mathbf{e}_{T}.
\end{align}
The distance vector from the center of RIS to the $i$-th reflective unit, denoted as $\mathbf{A}^{I}_{i}$,  is expressed as
\begin{align}
\mathbf{A}^{I}_{i}=\left(m_{i}-\frac{M_{I}+1}{2}\right)\Delta d_{y}^{I}\mathbf{e}^{c}+\left(n_{i}-\frac{N_{I}+1}{2}\right)\Delta d_{x}^{I}\mathbf{e}^{r},
\end{align}
where $m_{i}$ and $n_{i}$ represent the row number and column number of the $i$-th reflective unit.
Based on the above, the distances involving the specific antenna in Tx or the specific reflective unit in the RIS are given by
\begin{subequations}
\begin{align}
d^{TR}_{m}=\|\bm{\xi}_{R}-\bm{\xi}_{T}-\mathbf{A}^{T}_{m}\|,
\end{align}
\begin{align}
d^{TI}_{m,i}=\|\bm{\xi}_{I}-\bm{\xi}_{T}-\mathbf{A}^{T}_{m}+\mathbf{A}^{I}_{i}\|,
\end{align}
\begin{align}\label{dir_i}
d^{IR}_{i}=\|\bm{\xi}_{R}-\bm{\xi}_{I}-\mathbf{A}^{I}_{i}\|,
\end{align}
\begin{align}
d^{TS}_{m,c_{1},s} =\|\bm{\xi}_{c_{1},s}-\bm{\xi}_{T}-\mathbf{A}^{T}_{m}\|,
\end{align}
\begin{align}
d^{IS}_{i,c_{2},s}=\|\bm{\xi}_{c_{2},s}-\bm{\xi}_{I}-\mathbf{A}^{I}_{i}\|.
\end{align}
\end{subequations}

The angle parameters among Tx, Rx, RIS and clusters play an important role in radiation patterns. Based on the given 3-D coordinates and orientations of Tx, $\phi^{TR}$ can be computed directly as
$\phi^{TR}=\arccos\left(\frac{(\bm{\xi}_{R}-\bm{\xi}_{T})^{T}\mathbf{e}_{T}^{n}}{\|\bm{\xi}_{R}-\bm{\xi}_{T}\|}\right)$.
The calculations of  $\phi^{TI}$ and $\phi^{TC}_{c_{1}}$ are similar, thus omitted here. Based on the given 3-D coordinates and orientations of RIS, $\phi^{IT}$ is given by
$\phi^{IT}=\arccos\left(\frac{(\bm{\xi}_{T}-\bm{\xi}_{I})^{T}\mathbf{e}^{n}}{\|\bm{\xi}_{T}-\bm{\xi}_{I}\|}\right)$.
The calculations of  $\phi^{IR}$ and $\phi^{IC}_{c_{2}}$ are similar, thus omitted here. Based on the given 3-D coordinates and orientations of Rx, $\phi^{RT}$ is given by
$\phi^{RT}=\arccos\left(\frac{(\bm{\xi}_{T}-\bm{\xi}_{R})^{T}\mathbf{e}_{R}^{n}}{\|\bm{\xi}_{T}-\bm{\xi}_{R}\|}\right)$.
The calculations of $\phi^{RI}$, $\phi^{RC}_{c_{1}}$ and $\phi^{RC}_{c_{2}}$ are similar, thus omitted here.

After obtaining the distance parameters, the link attenuation can be calculated. The link attenuation is the linear gain caused by path loss. Let $PL^{\mathrm{LoS}}$ denote the LoS path loss in dB, and $L^{\mathrm{LoS}}=10^{\frac{-PL^{\mathrm{LoS}}}{10}}$. Numerous measurement activities have been conducted to fit the actual path loss in the complex indoor environment, resulting in various path loss models. Among them, the close-in free space reference distance  path loss model is applicable to various indoor environments including indoor hotspot (InH) and indoor factory (InF)  \cite{docomo20165g,jiang20213gpp,poddar2023tutorial}, thus applied here. Accordingly, we obtain
\begin{align}\label{pllos}
&PL^{\mathrm{LoS}}\\\nonumber
&=20\log_{10}\left(\frac{4\pi d_{0}}{\lambda}\right)+10n_{\mathrm{LoS}}\log_{10}\left(\frac{d^{TR}}{d_{0}}\right)+\chi_{\mathrm{LoS}},
\end{align}
where $d_{0}$ is the reference distance, typically set to be $1$ m, $n_{\mathrm{LoS}}$ represents the path loss exponent (PLE) of the LoS path,  $\chi_{\mathrm{LoS}}$ stands for the LoS shadow fading (SF) term (in dB), which is a zero-mean Gaussian random variable with a standard deviation  $\sigma^2_{\mathrm{LoS}}$. In Eq. (\ref{pllos}), the first term is the path loss under reference distances in the free space, the second term represents the path loss caused by the large propagation distance, and the third term is the path loss caused by the SF.

Let $PL^{\mathrm{VLoS}}$ denote the VLoS path loss in dB, and $L^{\mathrm{VLoS}}=10^{\frac{-PL^{\mathrm{VLoS}}}{10}}$. For the VLoS path, the traditional path loss model has been modified by considering two sub paths: from the Tx to the RIS and from the RIS to the Rx. Based on the modified model for RIS-assisted wireless propagation in \cite{Cui2025RIS,sang2023multi}, the VLoS path loss is given by
\begin{align}\label{plvlos}
&PL^{\mathrm{VLoS}}\\\nonumber
&=20\log_{10}\left(\frac{4\pi d^{1}_{0}d^{2}_{0}}{I\Delta d^{I}_{x}\Delta d^{I}_{y}  A}\right)+10n_{\mathrm{LoS}}\log_{10}\left(\frac{d^{TI}d^{IR}}{d^{1}_{0}d^{2}_{0}}\right)  \\\nonumber
&+\chi_{\mathrm{LoS}},
\end{align}
where  $d^{1}_{0}$ and $d^{2}_{0}$ denote the reference distances for $d^{TI}$ and $d^{IR}$ respectively, $A$ is the reflection magnitude of each reflective unit. In Eq. (\ref{plvlos}), the first term is the path loss under reference distances in free space resulting from the physical property of RIS \cite{tang2022path}, the second term represents the path loss caused by large propagation distance, and the third term is the path loss caused by SF. Note that, different from the path loss in   \cite{Cui2025RIS,sang2023multi}, the radiation pattern of reflective unit is extracted out of the path loss in this paper.

Let $PL_{c_{1}}^{\mathrm{SB-NLoS}}$ denote the SB-NLoS path loss involving the $c_{1}$-th cluster in dB, and  $L_{c_{1}}^{\mathrm{SB-NLoS}}=10^{\frac{-PL_{c_{1}}^{\mathrm{SB-NLoS}}}{10}}$. Based on the modeling of the LoS path, the SB-NLoS path loss involving the $c_{1}$-th cluster is given by
\begin{align}
&PL_{c_{1}}^{\mathrm{SB-NLoS}} \\\nonumber
&=20\log_{10}\left(\frac{4\pi d_{0}}{\lambda}\right)+10n_{\mathrm{NLoS}}\log_{10}\left(\frac{d^{TC}_{c_{1}}+d^{CR}_{c_{1}}}{d_{0}}\right) \\\nonumber
&+\chi_{\mathrm{NLoS}},
\end{align}
where $\chi_{\mathrm{NLoS}}$ stands for the NLoS SF term (in dB), which is a zero-mean Gaussian random variable with a standard deviation  $\sigma^2_{\mathrm{NLoS}}$, $d^{TC}_{c_{1}}$ and $d^{CR}_{c_{1}}$ represent the distance from Tx to the $c_{1}$-th cluster and the distance from the $c_{1}$-th cluster to Rx, respectively.

Let $PL_{c_{2}}^{\mathrm{DB-NLoS}}$ denote the DB-NLoS path loss involving the $c_{2}$-th cluster in dB, and  $L_{c_{2}}^{\mathrm{DB-NLoS}}=10^{\frac{-PL_{c_{2}}^{\mathrm{DB-NLoS}}}{10}}$. Based on the modeling of the VLoS path, the DB-NLoS path loss involving the $c_{2}$-th cluster is expressed as
\begin{align}
&PL_{c_{2}}^{\mathrm{DB-NLoS}}=20\log_{10}\left(\frac{4\pi d^{1}_{0}d^{2}_{0}}{I\Delta d^{I}_{x}\Delta d^{I}_{y}  A}\right) \\\nonumber
&+10n_{\mathrm{NLoS}}\log_{10}\left(\frac{d^{TI}(d^{IC}_{c_{2}}+d^{CR}_{c_{2}})}{d^{1}_{0}d^{2}_{0}}\right)+\chi_{\mathrm{NLoS}},
\end{align}
where $d^{IC}_{c_{2}}$ represents the distance from the RIS to the $c_{2}$-th cluster and  $d^{CR}_{c_{2}}$ represents the distances from the $c_{2}$-th cluster to Rx.

After obtaining the angle parameters, the radiation patterns can be calculated. In this paper, we do not limit the antenna types of Tx and Rx antennas, both directional and omnidirectional antennas can be selected according to the specific scenario. When an omnidirectional antenna is used, the radiation pattern is consistent across all directions. When a directional antenna is used, the radiation pattern is protruded around the main direction forming a main lobe.  In this paper, the cosine pattern, widely utilized in the antenna community \cite{balanis2016antenna,tang2022path}, is adopted for the directional antenna. Based on the above, when omnidirectional antennas are equipped on the Tx and Rx, the radiation patterns are expressed as $G_{T}\left( \phi\right)=1$,  $G_{R}\left( \phi\right)=1$ for any angle $\phi$. When directional antennas are equipped on the Tx and Rx, the radiation patterns of them are expressed as
\begin{subequations}
\begin{align}
G_{T}\left( \phi\right)= G_{T,\mathrm{max}}\cos^{\left(\frac{G_{T,\mathrm{max}}}{2}-1\right)}\phi,
\end{align}
\begin{align}
G_{R}\left( \phi\right)=G_{R,\mathrm{max}}\cos^{\left(\frac{G_{R,\mathrm{max}}}{2}-1\right)}\phi,
\end{align}
\end{subequations}
where $G_{T,\mathrm{max}}$ is the max gain of  Tx antenna while $G_{R,\mathrm{max}}$ is the max gain of Rx antenna. For the RIS, the cosine pattern is also widely used to express the radiation pattern of reflective unit \cite{tang2022path}, and it is given by
\begin{align}
G_{I}\left( \phi^{1}, \phi^{2}\right)=G_{I,\mathrm{max}}\cos\phi^{1}\cos\phi^{2},
\end{align}
where $G_{I,\mathrm{max}}$ is the max scattering gain of reflective unit, $\phi^{1}$ and $\phi^{2}$ are arbitrary incidence and reflection angles at RIS, respectively. According to the Appendix B of \cite{tang2022path}, $G_{I,\mathrm{max}}$  is a function of the surface area of the reflective unit, given by $G_{I,\mathrm{max}}=\frac{4\pi \Delta d_{x}^{I} \Delta d_{y}^{I}}{\lambda^2}$. This expression is also been verified through experimental measurements as shown in \cite{tang2022path,jamali2022low}.

\subsection{Statistical Cluster Generation Method}\label{scgmsection}
In the cluster-based model, the number of clusters, the number of scatterers in each cluster and the positions of these clusters and scatterers can be derived according to the given wireless application with actual measurements. To avoid a laborious measurement campaign, a statistical cluster generation procedure is often used. In this subsection, the statistical cluster generation procedure in \cite{basar2021indoor} is used for the SB-NLoS and DB-NLoS paths, but modified with more flexible settings.

Firstly, the clusters and scatterers in the SB-NLoS path are generated. The number of clusters typically follows a Poisson distribution \cite{basar2021indoor}, mathematically denoted by $C_{1} \sim \mathcal{P}(\lambda_{p})$, where $\lambda_{p}$ is a scenario-given constant.
The spatial distance from the Tx to the $c_{1}$-th cluster in the SB-NLoS path is modeled as a uniform random variable, i.e., $d^{TC}_{c_{1}} \sim \mathcal{U}[d_0, d^{TR}]$. With respect to the Tx broadside, the azimuth departure angle of the $c_{1}$-th cluster, denoted as $\phi^{SB}_{c_{1}}$, follows a $\mathcal{U}[\phi^{SB}_{b}, \phi^{SB}_{u}]$ distribution while the elevation departure angle of it, denoted as $\theta^{SB}_{c_{1}}$, follows a $\mathcal{U}[\theta^{SB}_{b}, \theta^{SB}_{u}]$ distribution.  $\phi^{SB}_{b}$, $\phi^{SB}_{u}$, $\theta^{SB}_{b}$ and $\theta^{SB}_{u}$ are the spread angle parameters of clusters with respect to the Tx  under a specific scenario.
In the $c_{1}$-th cluster, the number of scatterers follows a uniform distribution between the bounds $Sc_{b}$ and $Sc_{u}$, as verified for both the 28 GHz and 73 GHz frequency bands
\cite{hemadeh2017millimeter}. It is assumed that different scatterers within the $c_{1}$-th cluster have the same distance with respect to the Tx equal to $d^{TC}_{c_{1}}$, but have different azimuth/elevation departure angles with respect to the Tx \cite{basar2021indoor}. Within the $c_{1}$-th cluster, the azimuth departure angle of the $(c_{1},s)$-th scatterer, denoted by $\phi^{SB}_{c_{1},s}$, follows a conditionally Laplacian distribution with mean $\phi^{SB}_{c_{1}}$, i.e., $\phi^{SB}_{c_{1},s} \sim \mathcal{L}(\phi^{SB}_{c_{1}}, \sigma^{SB}_{\phi})$. Similarly, the elevation departure angle of the $(c_{1},s)$-th scatterer, denoted by $\theta^{SB}_{c_{1},s}$, satisfies $\theta^{SB}_{c_{1},s} \sim \mathcal{L}(\theta^{SB}_{c_{1}}, \sigma^{SB}_{\theta})$, where $\sigma^{SB}_{\phi}$ and $\sigma^{SB}_{\theta}$ are the standard deviations (angular spreads) of their Laplacian
distributions, which is given by the specific indoor environment \cite{basar2021indoor}. It should be noted that this modeling framework is inherently adaptable, allowing modifications of these small-scale fading parameters to suit various deployment environments.

As for the DB-NLoS path, which lies within the RIS-assisted region, the stochastic modeling framework applied to the SB-NLoS path is assumed to hold, with a key distinction being the reference point for distance and angular  definitions, now relative to the RIS broadside rather than the Tx broadside. The number of clusters $C_{2}$ is a Poisson random variable with $C_{2} \sim \mathcal{P}(\lambda_{p})$. With respect to the RIS, the spatial distance of the $c_{2}$-th cluster satisfies $d^{IC}_{c_{2}} \sim \mathcal{U}[d_0, d^{IR}]$, the azimuth departure angle, denoted as $\phi^{DB}_{c_{2}}$, follows a $\mathcal{U}[-\phi^{DB}_{b}, \phi^{DB}_{u}]$ distribution, and  the elevation departure angle, as $\theta^{DB}_c $, satisfies $\theta^{DB}_c \sim \mathcal{U}[-\theta^{DB}_{b}, \theta^{DB}_{u}]$ distribution. $\phi^{DB}_{b}$,  $\phi^{DB}_{u}$, $\theta^{DB}_{b}$ and $\theta^{DB}_{u}$ are the spread angle parameters of clusters with respect to the RIS.  In the $c_{2}$-th cluster, the number of scatterers follows a uniform distribution between the bounds $Sc_{b}$ and $Sc_{u}$. As for the $(c_{2},s)$-th scatterer,  the distance with respect to the RIS is equal to $d^{IC}_{c_{2}}$. Let $\phi^{DB}_{c_{2},s}$ and $\theta^{DB}_{c_{2},s}$  denote the azimuth departure angle and elevation departure angle, respectively. We have $\phi^{DB}_{c_{2},s} \sim \mathcal{L}(\phi^{DB}_{c_{2}}, \sigma^{DB}_{\phi})$ and $\theta^{DB}_{c_{2},s} \sim \mathcal{L}(\theta^{DB}_{c_{2}}, \sigma^{DB}_{\theta})$, where $\sigma^{DB}_{\phi}$ and $\sigma^{DB}_{\theta}$ are the standard deviations (angular spreads) of their Laplacian distributions.

When determining the distance and angle parameters with respect to the Tx and RIS, the positions of clusters and scatterers are naturally computed using Euclidean geometry.

\section{Spatial Consistency Procedure}\label{scpsection}
Using the channel modeling method proposed in Section \ref{cirsubsection}, one can generate the CIR of Rx at an arbitrary indoor position. Then the RSS at this position also can be computed.  According to the framework of RIS-WFL described in Section \ref{smpssection}, various indoor positions are surveyed where collecting fingerprint database, thus an overall indoor channels should be generated. It is known that the wireless propagation property remains nearly consistent between adjacent indoor positions. This implies that the spatial consistency should be considered when establishing the overall indoor channels.
In this section, spatial consistency procedures \cite{poddar2023tutorial} are used and extended to the RIS-assisted scenarios. Accordingly, the LoS/VLoS condition, the SF term, and the cluster-related parameters  are considered as the spatial consistent parameters.

For LoS/VLoS condition, a two-dimensional (2-D) grid map is generated. This map is designed to satisfy two requirements: the LoS/VLoS condition should follow a distance-dependent LoS/VLoS probability model at each location \cite{jiang20213gpp,docomo20165g}, while remaining spatially consistent \cite{poddar2023tutorial}. To achieve this, an initially independent Gaussian random field is generated on the grid and spatially filtered so that correlation decreases smoothly with distance. The correlated Gaussian field is then mapped to a unit-interval uniform field through a probability integral transform, which preserves the spatial correlation structure while providing a common probabilistic scale for decision-making. Finally, the LoS/VLoS condition at each grid  is obtained by comparing the uniform random value with the target LoS/VLoS probability predicted by the selected model at that location. This construction ensures both spatially coherent LoS/VLoS regions and correct marginal LoS/VLoS likelihoods consistent with the underlying probability model.

Before generating the 2-D grid map of LoS condition, a spatial consistency distance is defined as $d^{'}_{sc}$, then the granularity of the map is set to be $d^{'}_{sc}$. Let $(p,q)$ denote the index of the grid at the $p$-th row and $q$-th column, and the availability indicators of LoS path at this grid is denoted as $\zeta_{\mathrm{LoS},(p,q)}$. The steps of generating the grid map of LoS condition are given as follows:
\begin{enumerate}
  \item Each grid point is assigned an independent and identically distributed (i.i.d.) standard Gaussian random variable, generating an initial grid map;
  \item  The grid map is convolved with an exponential kernel to induce spatial correlation among the Gaussian random variables;
\item  A probability integral transform is applied to convert the Gaussian-distributed value at each grid to a uniform distribution over the interval [0,1],given by
\begin{equation}
u = \frac{1}{2} \left(1+\mathrm{erf}\left(\frac{v}{\sqrt{2}}\right)\right),
\end{equation}
where \( u \) and \( v \) are the uniform random variable and Gaussian random variable, respectively, and \( \mathrm{erf}(x) \) is the error function, given by $\mathrm{erf}(x) = \int_{-\infty}^{x} \frac{1}{\sqrt{2\pi}} e^{\frac{-t^2}{2}} dt$ for any $x$;
\item Compute the LoS probability of each grid according to a LoS probability model;
\item  The availability indicator is obtained by comparing the uniform random variable to the LoS probability at each grid, given by
\begin{equation}
\zeta_{\mathrm{LoS},(p,q)} = \left\{
\begin{aligned}
& 1  \quad \text{if} \quad u^{\mathrm{LoS}}_{p,q} \leq Pr_{\mathrm{LoS},(p,q)}(d^{TR}_{p,q}) \\
& 0  \quad \text{if} \quad u^{\mathrm{LoS}}_{p,q} > Pr_{\mathrm{LoS},(p,q)}(d^{TR}_{p,q})
\end{aligned}
\right.,~~\forall p,q.
\end{equation}
where $u^{\mathrm{LoS}}_{p,q}$ and $Pr_{\mathrm{LoS},(p,q)}$ are the uniform random variable and the LoS probability at the $(p,q)$-th grid,respectively, and $d^{TR}_{p,q}$ is the distance between the Tx and the $(p,q)$-th grid.
\end{enumerate}
The LoS probability model estimates the likelihood of direct visibility between the Tx and Rx according to their distance. Several LoS probability models have been proposed based on measurement campaigns, such as in 3GPP 38.901 model \cite{jiang20213gpp} and fifth generation channel model (5GCM) \cite{docomo20165g}.

\begin{figure*}
  \centering
  \includegraphics[width=0.9\textwidth]{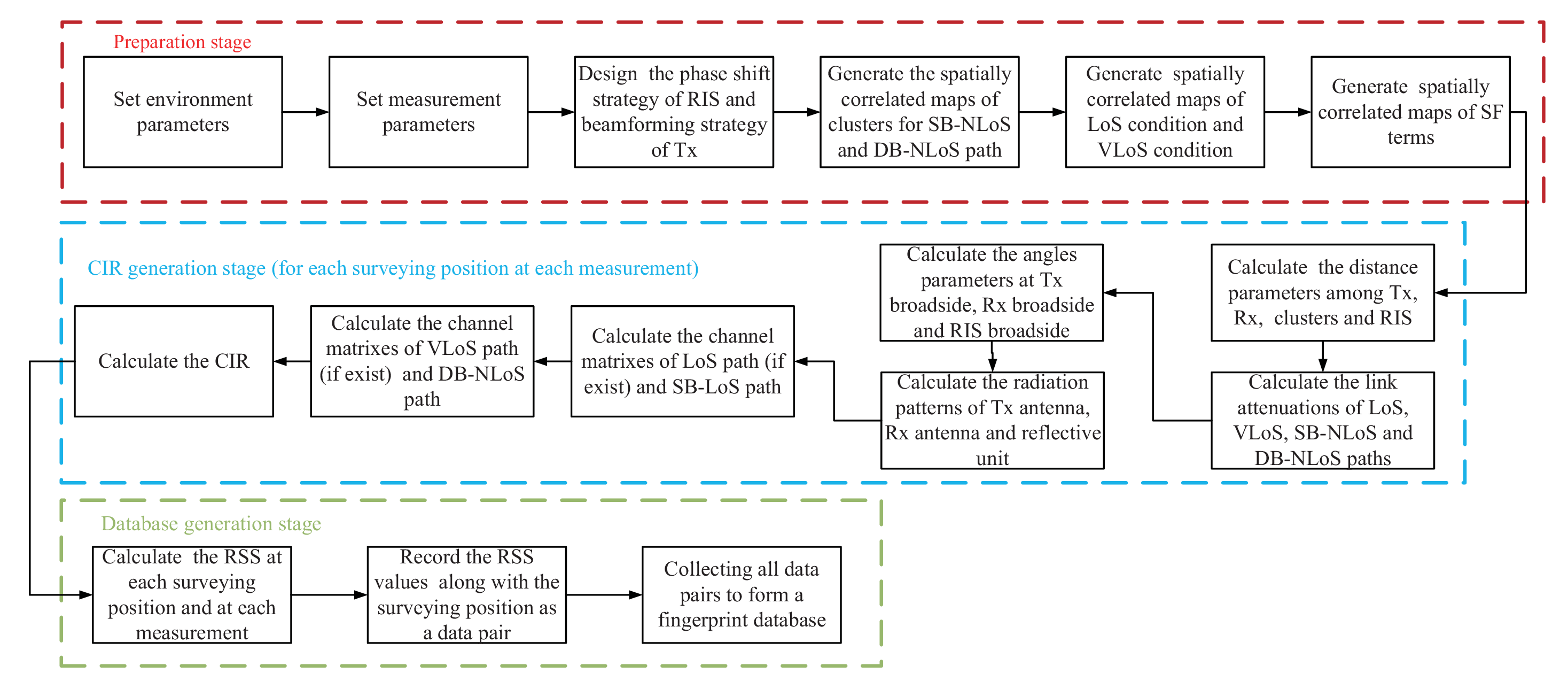}\\
  \caption{Framework of the proposed fingerprint database generation method.}\label{framfig}
\end{figure*}

Besides the LoS condition, the VLoS condition should be considered in the RIS-assisted scenario. Here, a 2-D grid map is also generated to contain values of the spatially correlated VLoS condition. The spatial consistency distance of VLoS condition is equal to $d^{'}_{sc}$. Thus, the granularity of this map is equal to $d^{'}_{sc}$. For the $(p,q)$-th grid, the availability indicator of VLoS path at this grid is denoted as $\zeta_{\mathrm{VLoS},(p,q)}$. Since in the same indoor environment,  the generation principle of the VLoS condition is assumed to be same as the LoS condition. Under the same approach to generate $u^{\mathrm{LoS}}_{p,q}$, at the ${p,q}$-th grid, a spatially  consistent uniformly distributed random variable is generated for the VLoS path, denoted as $u^{\mathrm{VLoS}}_{p,q}$. Then, the VLoS probability is calculated at each grid according to a VLoS probability model.
But different from the LoS case, when determining the VLoS condition, the relative geometric relationship between the RIS and the grid needs to be considered. This is because the RIS is a unidirectional surface, meaning it can only reflect electromagnetic waves into the space it faces. The reflective condition can be expressed based on the geometric relationship, and the availability indicator can be modified accordingly. Accordingly, the availability indicator of VLoS path, denoted as $\zeta_{\mathrm{VLoS},(p,q)}$, is given by
\begin{equation}
\zeta_{\mathrm{VLoS},(p,q)} = \left\{
\begin{aligned}
  & 1  \quad    u^{\mathrm{VLoS}}_{p,q} \leq Pr_{\mathrm{VLoS},(p,q)}(d^{IR}_{p,q}) \\
  & \quad  \cap (\bm{\xi}_{R,p,q}-\bm{\xi}_{I})^{T} \mathbf{e}^{n} > 0, \\
  & 0  \quad   u^{\mathrm{VLoS}}_{p,q} > Pr_{\mathrm{VLoS},(p,q)}(d^{IR}_{p,q}) \\
  & \quad  \cup (\bm{\xi}_{R,p,q}-\bm{\xi}_{I})^{T} \mathbf{e}^{n} \leq 0
\end{aligned}
\right.,
\end{equation}
where $\bm{\xi}_{R,p,q}$ is the position of the $(p,q)$-th grid,  $Pr_{\mathrm{VLoS},(p,q)}$ is the VLoS probability at the $(p,q)$-th grid, and $d^{IR}_{p,q}$ is the distance between the RIS and the $(p,q)$-th grid. Since in the same indoor environment, the form of the VLoS probability model is the same as the LoS probability model.

In SF modeling, spatially correlated random processes are used to simulate signal attenuation, as adjacent locations often experience similar fading. This process begins by generating a random field with a Gaussian distribution. The field is then filtered to introduce spatial dependencies, reflecting the effects of obstacles and terrain. Finally, the random field is standardized and scaled to match the desired statistical properties of the SF term, ensuring that the simulated signal attenuation aligns with prior conditions.

For the SF term of LoS path, a 2-D grid map is generated to contain spatially correlated values.  The spatial consistency distance of SF is defined as $d^{''}_{sc}$, then the granularity of this map is set to be $d^{''}_{sc}$. Let $\chi_{\mathrm{LoS},p,q}$ denote the SF term of LoS path at the ${p,q}$-th grid.
The steps of generating this grid map are given as follows:
\begin{enumerate}
  \item Assign an independent and identically distributed normal distributed random variable at each grid point, given by $\chi^{'}_{\mathrm{LoS},p,q}=\mathrm{randn}(0,\sigma^2_{LoS}),~\forall p,q$, where $\mathrm{randn}$ stands for the operation of generating a Gaussian random number.
  \item A 2-D exponential filter is established, whose filter response at the $(p,q)$-th grid is given by the following equation;
\begin{align}
h(p,q)=e^{-\frac{\sqrt{p^2+q^2}}{d^{\mathrm{SF}}_{co}}},
\end{align}
where $d^{\mathrm{SF}}_{co}$ is the correlation distance of SF;
\item Calculate the correlated values in this grid map, given by
\begin{align}
\chi^{'}_{\mathrm{LoS},p,q}=\sum_{i} \sum_{j} h(i,j) \chi_{\mathrm{LoS},p-i+1, q-j+1}.
\end{align}
\item Compute the mean and standard deviation of $\chi^{'}_{\mathrm{LoS}}$, denoted as $\mu^{'}_{\mathrm{LoS}}$ and $\sigma^{'}_{\mathrm{LoS}}$, respectively. Scale the grid map to match the statistical properties of SF term, given by
\begin{align}
\chi_{\mathrm{LoS},p,q}=\chi^{'}_{\mathrm{LoS},p,q}\frac{\sigma_{\mathrm{LoS}}}{\sigma^{'}_{\mathrm{LoS}}}-\mu^{'}_{\mathrm{LoS}}.
\end{align}
\end{enumerate}

For the SF term of NLoS path, a 2-D grid map is also generated to contain spatially correlated values. The granularity of this map is equal to $d^{''}_{sc}$. Let $\chi_{\mathrm{NLoS},p,q}$ denote the SF term of NLoS path at the ${p,q}$-th grid. This grid map is generated following the same steps in the LoS case, thus omitted here. The only difference is that the standard deviation of the SF term is $\sigma_{\mathrm{NLoS}}$, rather than $\sigma_{\mathrm{LoS}}$.

Finally, the spatially correlated map of clusters is generated.
It is reasonable to assume that the same clusters and scatterers are shared within a coherence distance. Let $d^{'''}_{sc}$ denote the coherence distance. Then, the granularity of this map is set to $d^{'''}_{sc}$. For each grid, the amount and positions of clusters and scatterers are generated according to the statistical generation method shown in Section \ref{scgmsection}, independently.

\section{Fingerprint Database Generation Method}\label{secfdgf}

In this section, the fingerprint database generation method is proposed based on the CIR modeling method and the spatial consistency procedure.

Before illustrating the method for generating the database, we first introduce the relationship between the fingerprint database and the CIRs. Let $\mathcal{P}$ denote the set of surveying positions where fingerprint data are collected, as illustrated in Section.~\ref{smpssection}. Let $P$ denote the number of surveying positions. At each surveying position, assume $N$ measurements are taken at Rx. At the $n$-th measurement, the beamforming of Tx is configured as $\mathbf{f}^{n}_{T}$, and the phase shift matrix of RIS is configured as $\bm{\Omega}^{n}$. Let $\mathbf{H}^{n}_{p}(\tau)$ \footnote{Note that $\mathbf{H}^{n}_{p}(\tau)$ is a function of $\bm{\Omega}^{n}$, since the VLoS path and DB-NLoS path are related to the phase shift matrix, as shown in Eq.~(\ref{channelvlos}) and Eq.~(\ref{channeldbnlos}).} and $\mathrm{RSS}^{n}_{p}$  denote the CIR and the RSS at the $p$-th surveying position at the $n$-th measurement, respectively\footnote{In this paper, RSS is selected as the wireless fingerprint. However, our approach can be easily extended  to a channel state information (CSI)-based WFL system, as the CSI matrix can be directly computed using the CIRs \cite{xu2024swin}.}. We have
\begin{align}\label{rsscomputeeq}
\mathrm{RSS}^{n}_{p}=10\log_{10}\left(\left|\sum_{\tau=0}^{\tau_{\mathrm{max}}} \sqrt{P_{0}} \mathbf{H}^{n}_{p}(\tau)\mathbf{f}_{t}^{n} x^{n}\right|^2\right),
\end{align}
where $P_{0}$ is the transmit power of Tx, and $x^{n}$ is the transmit symbol, which is usually set to be 1. At the $p$-th surveying position, the fingerprint data pair is given by  $(\{\mathrm{RSS}^{n}_{p}\}_{n=1}^{N}, \mathcal{P}(p))$, where $\mathcal{P}(p)$ denotes the coordinate of this position. By collecting all fingerprint data pairs, the fingerprint database, denoted as $\mathcal{D}$, is given by
\begin{align}\label{databasegeeq}
\mathcal{D}=\{\{\mathrm{RSS}^{n}_{p}\}_{n=1}^{N}, \mathcal{P}(p)\}_{p=1}^{P}.
\end{align}

The proposed fingerprint database generation method is outlined in Fig.~\ref{framfig}. It consists of the preparation stage, CIR generation stage and database generation stage. The preparation stage establishes basic settings and spatially consistent maps of channel parameters. The CIR generation stage establishes the CIRs at surveying positions, and the database generation stage establishes the fingerprint database.

The steps of the preparation stage in the proposed method are outlined  as follows:
\begin{enumerate}
  \item Set environment parameters including electromagnetic parameters of Tx, Rx and RIS, positions and orientations of them and wireless propagation parameters;
  \item Set measurement parameters including the number of measurement and surveying positions;
   \item Design beamforming strategy of Tx, and phase shift strategy of RIS, namely determine the beamforming and phase shift configurations at different measurement;
  \item Generate the spatially correlated maps of clusters for the SB-NLoS path: Determine the number of clusters $C_{1}$  and the number of scatterers $S_{c_{1}}$, for $\forall c_{1}$. Generate the azimuth departure angle $\phi^{SB}_{c_{1},s}$ and the elevation departure angle  $\theta^{SB}_{c_{1},s}$ for $\forall c_{1}, \forall s$. Generate the distances between the Tx and the clusters $d^{TC}_{c_{1}}$, for $\forall c_{1}$. Calculate the positions of the clusters and internal scatterers in the SB-NLoS path. Generate the complex path gain resulted from each scatterer;
\item  Generate the spatially correlated maps of clusters for the DB-NLoS path: Determine the number of clusters $C_{2}$  and sub-rays  $S_{c_{2}}$, for $\forall c_{2}$. Generate the azimuth departure angle $\phi^{DB}_{c_{2},s}$ and the elevation departure angle $\theta^{DB}_{c_{2},s}$, for $\forall c_{2}, \forall s$. Generate the distances between the RIS and the clusters $d^{IC}_{c_{2}}$, for $\forall c_{2}$. Calculate the positions of the clusters and internal scatterers in the DB-NLoS path. Compute the complex path gains resulted from each scatterer;
\item Generate the spatially correlated maps of LoS and VLoS conditions:  Generate spatially correlated uniform random variables,  and generate the LOS probabilities. Then determine the availability indicators of LoS/VLoS path by comparing the uniform random variable to the LoS/VLoS probability;
\item Generate the spatially correlated SF maps: Generate independent and identically distributed normal distributed random variables. Then generate a spatially correlated map of SF in the LoS/NLoS case by a 2-D exponential filter.
\end{enumerate}

In the CIR generation stage, $N$ CIRs are generated at each surveying position under $N$ measurements. At each surveying position and for each measurement, the CIR is generated by following steps:
\begin{enumerate}
\item Calculate the distance parameters between the Tx and Rx, the distance parameters between the Tx and RIS,  the distance parameters between the RIS and Rx, the distance parameters between the Tx and clusters/scatterers in the SB-NLoS path, the distance parameters between the clusters/scatterers and Rx in the SB-NLoS path, the distance parameters between the RIS and clusters/scatterers in the DB-NLoS path, and the distance parameters between the clusters/scatterers and Rx in the DB-NLoS path;

\item Based on the spatially correlated maps of shadow fading and distance parameters, calculate the link attenuations, i.e., $L^{\mathrm{LoS}}$, $L^{\mathrm{VLoS}}$, $L^{\mathrm{SB-NLoS}}$ and $L^{\mathrm{DB-NLoS}}$;

\item Calculate the angles of departure at the Tx broadside, i.e., $\phi^{TR}$, $\phi^{TI}$ and $\phi^{TC}_{c_{1}}$.  Calculate the angles of arrival and the angles of departure at RIS broadside, i.e., $\phi^{IT}$, $\phi^{IR}$ and $\phi^{IC}_{c_{2}}$. Calculate the angles of arrival at Rx broadside, i.e., $\phi^{RT}$, $\phi^{RI}$, $\phi^{RC}_{c_{1}}$ and $\phi^{RC}_{c_{2}}$;

\item Based on the antenna and RIS property and angle parameters, calculate the radiation patterns of Tx antenna, Rx antenna and reflective unit;

\item Combining the spatially correlated map of LoS condition, calculate the channel matrix of LoS path according to Eq.~(\ref{channellos}) (if the LoS condition is true) , and calculate the channel matrix of SB-NLoS path according to Eq.~(\ref{channelsbnlos});

\item Combining the spatially correlated map of VLoS condition and the phase shift configuration of RIS, calculate the channel matrix of VLoS path according to Eq.~(\ref{channelvlos}) (if the VLoS condition is true), and calculate the channel matrix of DB-NLoS path according to Eq.~(\ref{channeldbnlos});
\item Generate the CIR according to Eq.~(\ref{totallchannel}).
\end{enumerate}

Finally, the steps of the database generation stage are outlined as follows:
\begin{enumerate}
\item Based on the CIR and the beamforming configuration of Tx, calculate the RSS values at each surveying position and each measurement according to Eq.~(\ref{rsscomputeeq});
\item Record the RSS values with multiple measurements along with the surveying positions of Rx to form fingerprint data pairs;
\item Generate the fingerprint database $\mathcal{D}$ by collecting all data pairs according to Eq.~(\ref{databasegeeq}).
\end{enumerate}

\section{Extension and Discussion}
In this section, we resolve  synchronization and RIS phase configurations setting issues for the practical application of the RIS-assisted WFL illustrated in Sec.\ref{smpssection}, and also extend the database generation method with MC effect.
\subsection{Synchronization}\label{secsyn}
In the offline phase, the measurement activity is controllable. When the device reaches a specific grid, the network operator reconfigures the RIS and changes the RIS configurations with a predetermined order in a time-slot sequence. The number of time slots is just equal to the number of measurements $N$. Under these settings, the RSS values collected by the device follow a predetermined order. However, in the online phase, there are multiple third-party devices requiring localization, it is not possible to reconfigure the RIS for every device. Alternatively, the RIS configurations are set periodically. But, in this way, the measured $N$ RSS values of one online device are disordered, as the first measurement may not be within the predetermined first time slot.

To solve the above synchronization issue, we propose a simple and effective sorting algorithm. In the online phase, the device sends the measured RSS value sequence, denoted as $\mathbf{RSS}^{'}_{0}$, to the network operator, which then sorts the sequence using the proposed algorithm to match the offline order. First, all possible sequences with different starting time slots are generated from this sequence via the circular shift. In this way, a sequence set $\{\mathbf{RSS}^{'}_{n}\}_{n=0}^{N-1}$ is obtained. The $n$-th sequence ($n=1,2,\cdots, N-1$) in the set is given by
\begin{align}\label{rss'n}
&\mathbf{RSS}^{'}_{n}=\\\nonumber
&\begin{bmatrix} \mathbf{RSS}^{'}_{n-1}(N) \mathbf{RSS}^{'}_{n-1}(1)  \mathbf{RSS}^{'}_{n-1}(2)\cdots \mathbf{RSS}^{'}_{n-1}(N-1)\end{bmatrix}. \\\nonumber
\end{align}
Then, for each RSS value sequence in the set, compute the distance vector between this sequence and the offline database, shown in Eq.~(\ref{databasegeeq}). For the $n$-th RSS value sequence, the distance vector, denoted as $\mathbf{d}^{\mathrm{syn}}_{n}$, is given by
\begin{align}\label{rss'n}
\mathbf{d}^{\mathrm{syn}}_{n}(p)=\|\mathbf{RSS}^{'}_{n}-\mathbf{RSS}_{p}\|, ~p=1,2,P,
\end{align}
where $\mathbf{RSS}_{p}$ is the $p$-th RSS value sequence in the offline database, given by
 $\mathbf{RSS}_{p}=\begin{bmatrix} \mathrm{RSS}^{1}_{p} & \mathrm{RSS}^{2}_{p} & \cdots & \mathrm{RSS}^{N}_{p} \end{bmatrix}$. Then, the min value in $\mathbf{d}^{\mathrm{syn}}_{n}$ is recorded, denoted as $E^{\mathrm{min}}_{n}$. Finally, find the index of min value in $\{E^{\mathrm{min}}_{n}\}_{n=0}^{N-1}$, expressed as
\begin{align}\label{databasegeeq}
n^{*}=\underset{n}{\arg\min}~ E^{\mathrm{min}}_{n}, ~n=0,1,\cdots,N-1.
\end{align}
Finally, take $\mathbf{RSS}^{'}_{n^{*}}$ as the ordered RSS value sequence, which is synchronous with the offline phase.

\subsection{RIS Phase Configuration Strategy}
The wireless fingerprint in the RIS-assisted WFL refers to RSS values collected from multiple measurements under different RIS phase configurations. Distinct RIS phase configurations yield different radio maps. As the diversity of radio maps increases, the separability of fingerprints across spatial locations improves, thereby enhancing localization accuracy. However, because a typical RIS contains hundreds or even thousands of reflecting elements, the space of possible configurations is immense, making it nontrivial to select optimal configurations for multiple measurements.

Here, we propose an empirical RIS phase configuration strategy, named uniform focal scanning  strategy (UFSS). In UFSS, within each measurement, the RIS phase shifts are set according to the rate-maximization scheme \cite{cheng2021joint}, so that signal energy from the VLoS path is focused at a particular position in the AoI. And across multiple measurements, these focal positions are uniformly distributed over the AoI. By shifting the focal position of the reflected energy, UFSS generates largely distinct radio maps and thereby increases the spatial resolution of the fingerprint database.

UFSS consists of the following steps. Firstly, the AoI is partitioned into \(N\) uniform grids, and the center of the \(n\)-th grid, is selected as the desired focal position of the reflected energy during the \(n\)-th measurement. Then the diagonal phase shift matrix at the $n$-th measurement is given by
\begin{align}
\bm{\Omega}^{n}(i,i)
= e^{ j\frac{2\pi}{\lambda}\Big(\big\|\bm{\xi}_{I}-\bm{\xi}_{T}+\mathbf{A}^{I}_{i}\big\|+\big\|\bm{\xi}_{R,n}-\bm{\xi}_{I}-\mathbf{A}^{I}_{i}\big\|\Big) },
\end{align}
where $\bm{\Omega}^{n}(i,i)$ is the $i$-th diagonal element in $\bm{\Omega}^{n}$, $\bm{\xi}_{R,n}$ is the position of the \(n\)-th grid. In this way, the RIS phase configurations for all \(N\) measurements are determined.
\subsection{RIS Phase Profile with MC}\label{sectionmodelmc}
\begin{figure}
  \centering
  \includegraphics[width=0.5\textwidth]{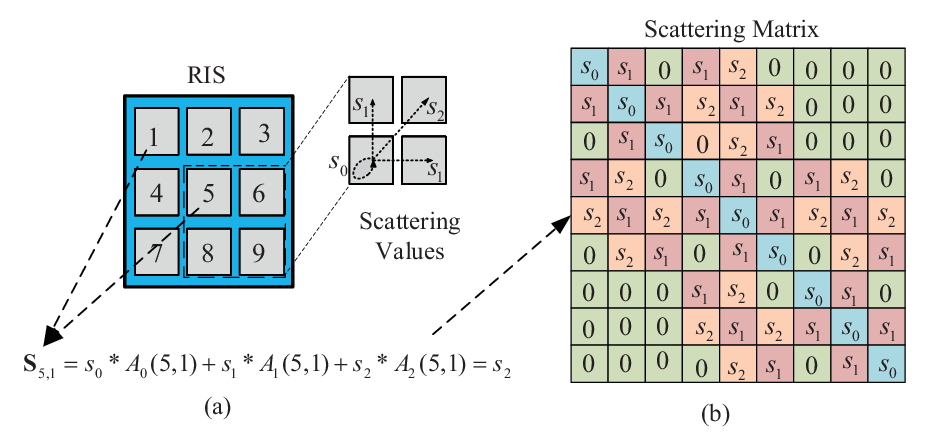}\\
  \caption{Illustration of scattering values and the scattering matrix when $B_{M}=2$ and $M_{I}=N_{I}=3$. (a) Scattering values. (b) The scattering matrix.}\label{couplingillufig}
\end{figure}

In Sec. \ref{cirsubsection}, we define the phase shift matrix of RIS as a diagonal matrix whose diagonal elements represent the phase shifts of  reflective units. But in some scenarios where MC can not be ignored, the modeling should be adjusted \cite{fadakar2025mutual,zheng2024mutual,Fadakar2025NearFieldRL}. The MC arises from the  interaction between adjacent reflective units, where the EM fields generated by one unit affect the behavior of nearby units \cite{fadakar2025mutual}. Based on the microwave network theory, the scattering matrix-based model is proposed to represent the EM interactions of reflective units in RIS, and the scattering matrix directly functions on the ideal diagonal phase shift matrix of RIS \cite{zheng2024mutual}. Let $\bm{\Omega}_{\mathrm{MC}}$ denote the phase shift matrix with MC effect. The relationship of it and the ideal phase shift matrix $\bm{\Omega}$ is given by
\begin{align}\label{mcformula}
\bm{\Omega}_{\mathrm{MC}}=(\bm{\Omega}^{-1}-\mathbf{S})^{-1},
\end{align}
where $\mathbf{S}$ denote the scattering matrix.

Specifically, the entry $\mathbf{S}(i,j)$ in $\mathbf{S}$ is the scattering value between the $i$-th reflective unit and the $j$-th reflective unit on the RIS, which indicates the voltage wave measured at the $i$-th unit when a unit voltage wave is driven at the j-th unit \cite{zheng2024mutual}. Typically, the EM interaction decreases with increasing distance between unit pairs, so only the scattering values of unit pairs with row and column index differences within \( B_M \) are considered \cite{zheng2024mutual}. Based on this, the number of unique scattering values that each reflective unit can influence, denoted as \( K \), is given by  $K=\frac{(B_M+1)B_M}{2}$. And the \( k \)-th scattering value is denoted as \(s_k\), where \(k = 0, 1, 2, \dots, K-1 \). Based on the above, the scattering matrix is given by
\begin{align}\label{defispara}
\mathbf{S}=\sum_{k=0}^{K-1} s_{k}\mathbf{A}_{k},
\end{align}
where $\mathbf{A}_{k}\in \mathbb{Z}^{I\times I}$ accounts for the support matrix of  $s_{k}$, whose units are binary variables. $\mathbf{A}_{k}(p,q)=1$ indicates that the voltage wave $s_{k}$ is measured at the $p$-th reflective unit when a unit voltage wave is applied at the $q$-th reflective unit.

Assume that the reflective units exhibit rotation-symmetric, and the physical conditions such as temperature of all units are identical, then the element of the support matrix only relates to the row index and column index \cite{zheng2024mutual}. Here, we propose a general expression of the support matrix, given by
\begin{subequations}\label{supportexpression}
\begin{align}
\mathbf{A}_{k}(p,q)=\begin{cases}
1 &  F \\
0 &  \overline{F}
\end{cases},
\end{align}
\begin{align}
F:&\left(|r_p-r_q|+|c_p-c_q|=k\right)
 \cap \left(|r_p-r_q|\leq B_{M}-1\right) \\\nonumber
 &\cap \left(|c_p-c_q|\leq B_{M}-1\right),
\end{align}
\end{subequations}
where $\overline{F}$ is the opposite condition of $F$, $r_p$ and $r_q$ represent the rows of the $p$-th and the $q$-th unit in RIS while $c_p$ and $c_q$ represent the columns of the $p$-th and the $q$-th unit in RIS. In Eq.~(\ref{supportexpression}), $\left(|r_p-r_q|\leq B_{M}-1 \cap |c_p-c_q|\leq B_{M}-1\right)$ results form the limit of the EM interaction range, and $|r_p-r_q|+|c_p-c_q|=k$ matches the definition of the $k$-th scatter value. For ease of understanding, the scattering matrix for $B_{M}=2$ and $M_{I}=N_{I}=3$ is illustrated in Fig.~\ref{couplingillufig}.

For a specific scenario, the scattering values are obtained through measurement or estimation, and the support matrix is computed using Eq.~(\ref{supportexpression}). Then, the scatter matrix is computed using Eq.~(\ref{defispara}). Finally, the phase shift matrix of RIS with MC is computed using Eq.~(\ref{mcformula}).

\section{Simulation Results}

\subsection{Simulation Setup}
Here, we consider a general indoor environment with size $20\times20\times3.5 \mathrm{m}^3$, and its environment parameters are referred from InH \cite{poddar2023tutorial,3GPP_TR_38_901}.
The large-scale environment parameters are set as follows: $n_{\mathrm{LoS}}=1.73$, $\sigma_{\mathrm{LoS}}=3.02$, $n_{\mathrm{NLoS}}=3.19$, $\sigma_{\mathrm{NLoS}}=8.29$, $\sigma^{SB}_{U}=\sigma^{DB}_{U}=6$,$d^{'}_{sc}=1 \mathrm{m}$, $d^{''}_{sc}=2 \mathrm{m}$, $d^{'''}_{sc}=2.5 \mathrm{m}$, $d^{\mathrm{SF}}_{co}=4 \mathrm{m}$. The small-scale environment parameters for generating clusters and scatterers are set as follows: $\lambda_{p}=1.8$, $Sc_{b}=1$, $Sc_{u}=30$, $\sigma^{SB}_{\phi}=\sigma^{DB}_{\phi}=5^{o}$,
$\sigma^{SB}_{\theta}=\sigma^{DB}_{\theta}=5^{o}$, $\phi^{SB}_{b}=-90^{o}$, $\phi^{SB}_{u}=90^{o}$,
$\theta^{SB}_{b}=-45^{o}$, $\theta^{SB}_{u}=45^{o}$, $\phi^{DB}_{b}=225^{o}$, $\phi^{DB}_{u}=315^{o}$,
$\theta^{DB}_{b}=-45^{o}$, $\theta^{DB}_{u}=45^{o}$. The frequency of wireless signal is 5.2 GHz.

A local coordinate system is established, where the x-coordinate spans the length, the y-coordinate spans the width, and the z-coordinate spans the height of the indoor environment, as shown in Fig.~\ref{riswtfig}. A Tx is placed in the y-z plane, and its transmit power is $10~ \mathrm{dBm}$. A ULA is equipped at Tx with $\frac{\lambda}{2}$ antenna spacing. The Tx antenna is directional with an 8 dB max gain. The default number of antennas is set to $4$. A RIS is placed in the x-z plane. The default number of reflective units in RIS is set to $400$ while the default row number and column number of reflective unit array are both set as 20. The length and width of each reflective unit are both equal to $\frac{\lambda}{2}$. The positions of Tx and RIS are given by $\bm{\xi}_T=[0, 10, 3] (\mathrm{m})$ and $\bm{\xi}_I=[10, 15, 3] (\mathrm{m})$ while the orientations are given by $\mathbf{e}_T=[0,0,-1]$,  $\mathbf{e}_T^n=[1,0,0]$, $\mathbf{e}^{r}=[0,0,1]$, $\mathbf{e}^{c}=[1,0,0]$ and $\mathbf{e}^{n}=[0,-1,0]$. A Rx equipped with an omnidirectional antenna is movable with height $1 \mathrm{m}$, and the orientation of Rx is given by $\mathbf{e}_{R}^{n}=[0,0,1]$.

The AoI for the positioning server is a subarea of the above indoor environment with length 10 m (from 5 m to 15 m on the x axis) and width 10 m (from 0 m to 10 m on the y axis)\footnote{For the sake of tractability, a 2-D AoI is used in simulations. When a 3-D AoI is considered, only the coordinates of Rx are modified from 2-D vectors to 3-D vectors to compute distances and angles, while the rest of the simulation process remains unchanged.}.  To develop a fingerprinting database, the AoI is divided into a grid with uniform spacing of $0.2 \mathrm{m}$ in each dimension. At each grid center, RSS values are measured multiple times. The default number of measurements is $20$. By default, the proposed UFSS is used to determine the RIS phase configurations. With these configurations, fingerprint databases are generated.

To evaluate the RIS-assisted WFL, the typical convolutional neural network (CNN) algorithm \cite{ibrahim2018cnn}, KNN algorithm \cite{wu2023fingerprint,sardellitti2024ris,hou2025indoor}, DNN algorithm \cite{zhang2022multiple} and Gaussian process (GP) algorithm \cite{kumar2016gaussian} are employed.  The CNN includes two convolutional layers with one-dimensional convolution kernels, followed by two fully connected layers, with each layer applying ReLU activations, and concluding with a regression layer for the output. For the DNN, a four-layer network is established, the first three layers of it apply the hyperbolic tangent sigmoid function while the last layer of it apply the linear activation function. More details of these algorithms can be found in the open-source code\footnote{The database, proposed generation method and localization algorithms are publicly available at https://github.com/XinCheng2025/RIS-WFL-DB-Generation-Method.\label{footnotecode}}. When evaluating the localization performance, the fingerprint database is randomly divided into a training database and a test database with an 8:2 ratio.

\begin{figure}
   \subfloat[ \label{LoSmap1}Map of LoS condition.]{%
       \includegraphics[width=0.24\textwidth]{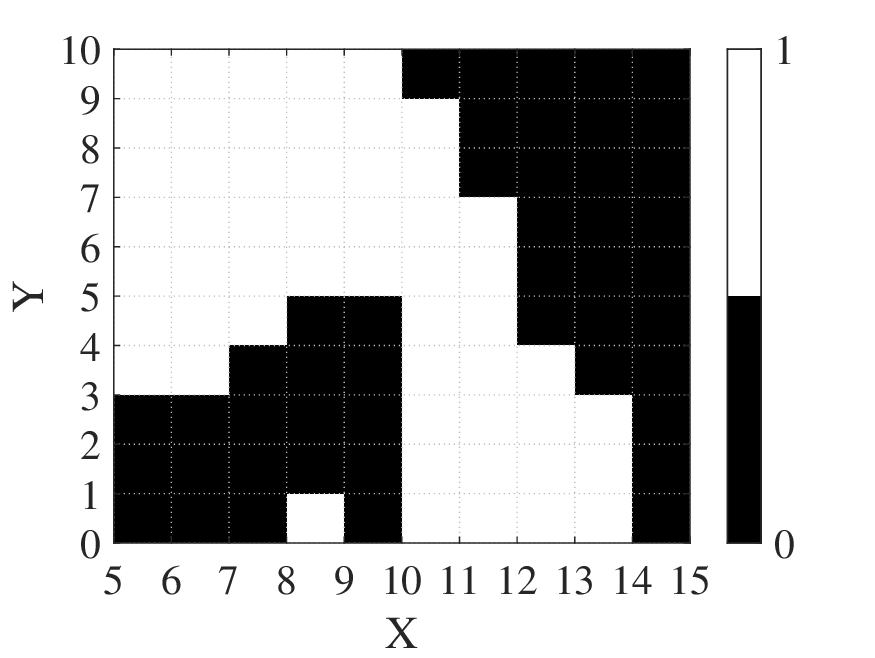}
     }
      \subfloat[ \label{VLoSmap2}Map of VLoS condition.]{%
       \includegraphics[width=0.24\textwidth]{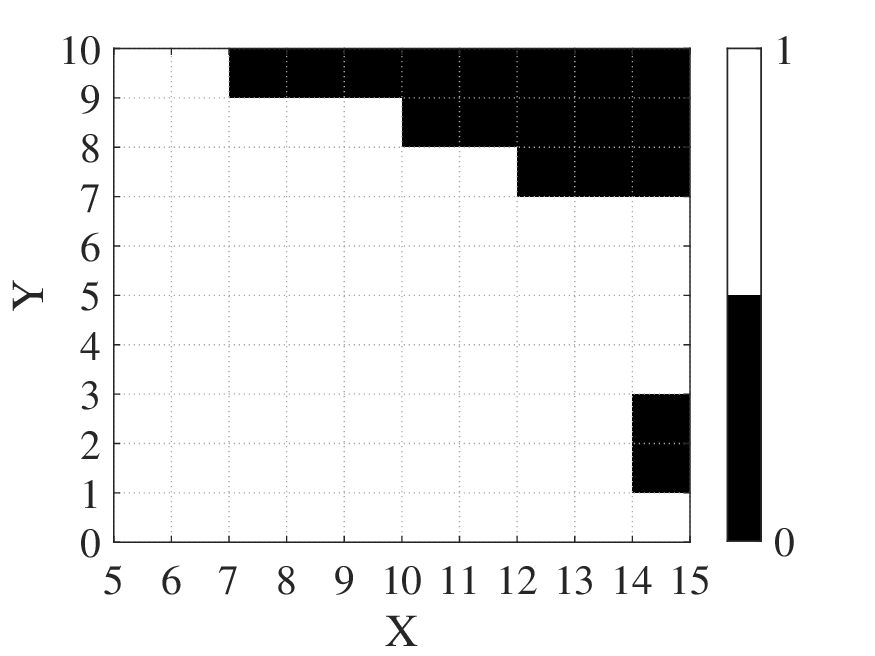}
     }
     \caption{Spatially correlated maps of LoS and VLoS conditions (1 indicates the LoS/VLoS path is available while 0 indicates the LoS/VLoS path is blocked).}\label{losmapfig}
\end{figure}
\begin{figure}
   \subfloat[ \label{LoSmap1}Map of SF in LoS paths.]{%
       \includegraphics[width=0.24\textwidth]{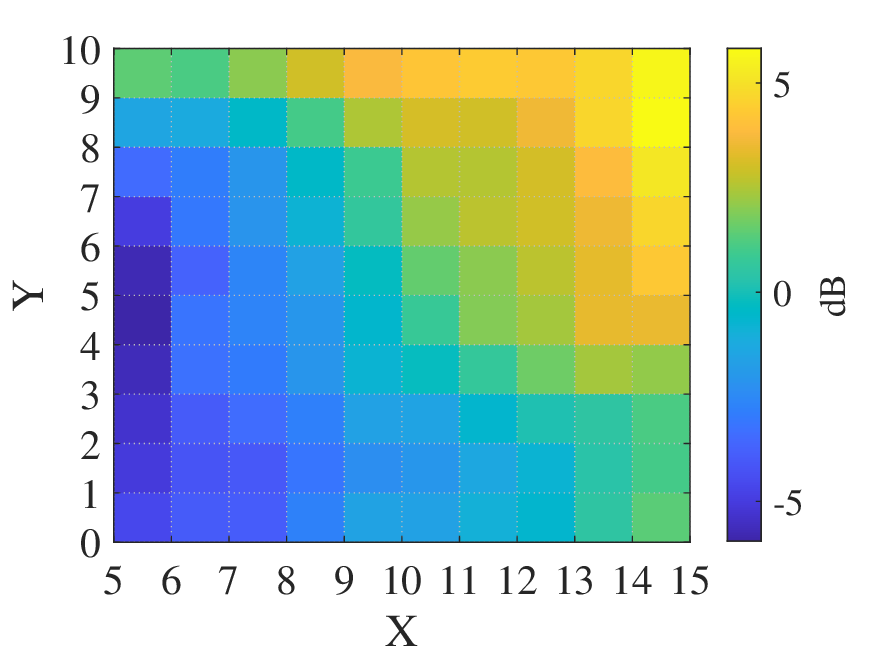}
     }
      \subfloat[\label{VLoSmap2}Map of SF in NLoS paths.]{%
       \includegraphics[width=0.24\textwidth]{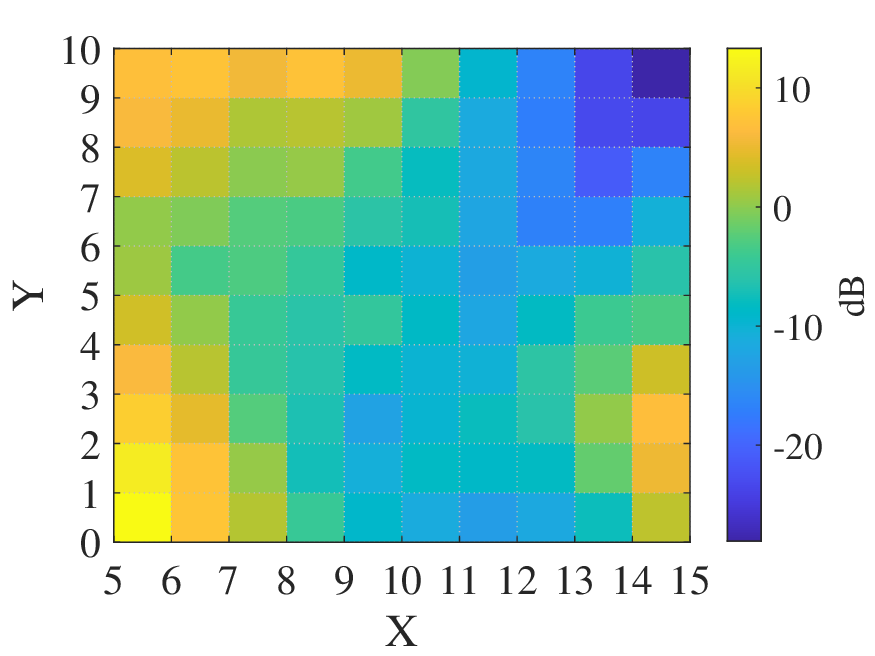}
     }
     \caption{Spatially correlated SF maps in LoS and NLoS paths.}\label{sfmapfig}
\end{figure}

\subsection{Property of Generated Fingerprint Database }
Using the proposed method, we generate an open-source fingerprint database named RIS-WFL-DB\textsuperscript{\ref{footnotecode}} to facilitate related research. The spatially correlated maps of LoS condition and VLoS condition when generating RIS-WFL-DB are shown in Fig.~\ref{losmapfig}. As expected, adjacent positions in the AoI usually share the same LoS or VLoS condition. The maps of SF in LoS paths (the LoS path and VLoS path) and NLoS paths (the SB-NLoS path and DB-NLoS path) when generating RIS-WFL-DB are illustrated in Fig.~\ref{sfmapfig}. It shows smooth variations of SF across adjacent positions in the AoI.

\begin{figure*}
   \subfloat[ \label{rss1}At the 3-th measurement.]{%
       \includegraphics[width=0.24\textwidth]{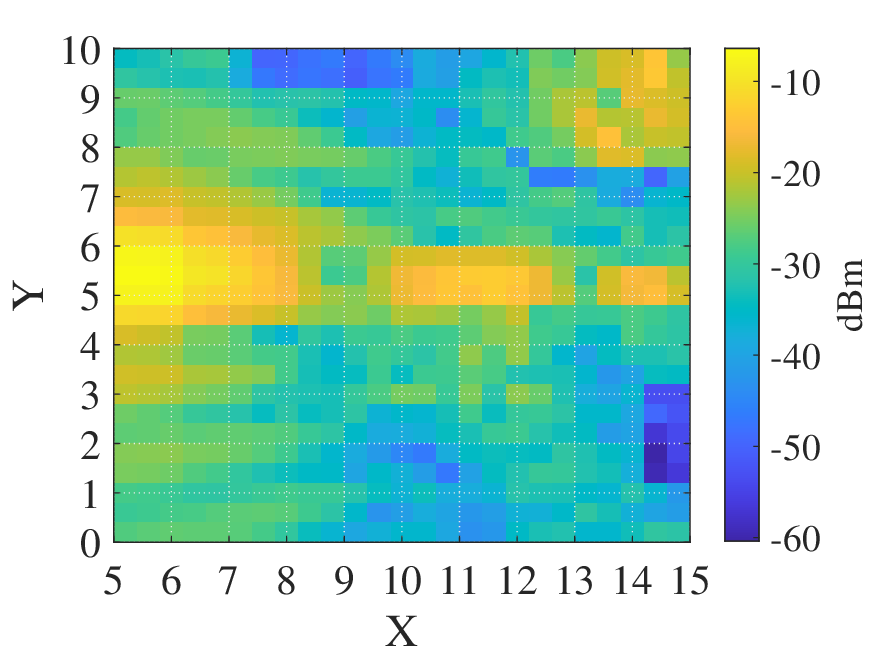}
     }
      \subfloat[ \label{rss2}At the 7-th measurement.]{%
       \includegraphics[width=0.24\textwidth]{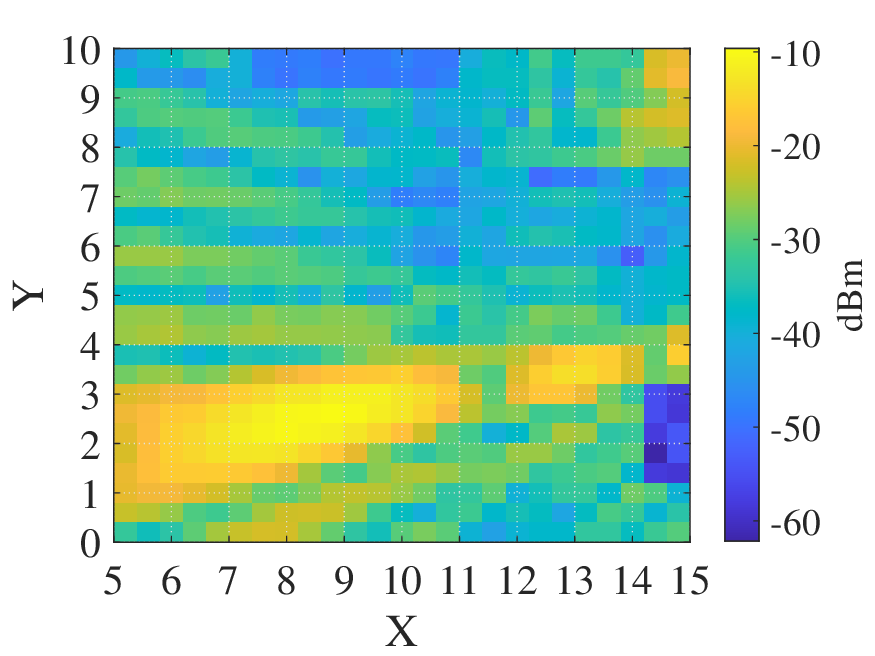}
     }
    \subfloat[ \label{rss3}At the 10-th measurement.]{%
       \includegraphics[width=0.24\textwidth]{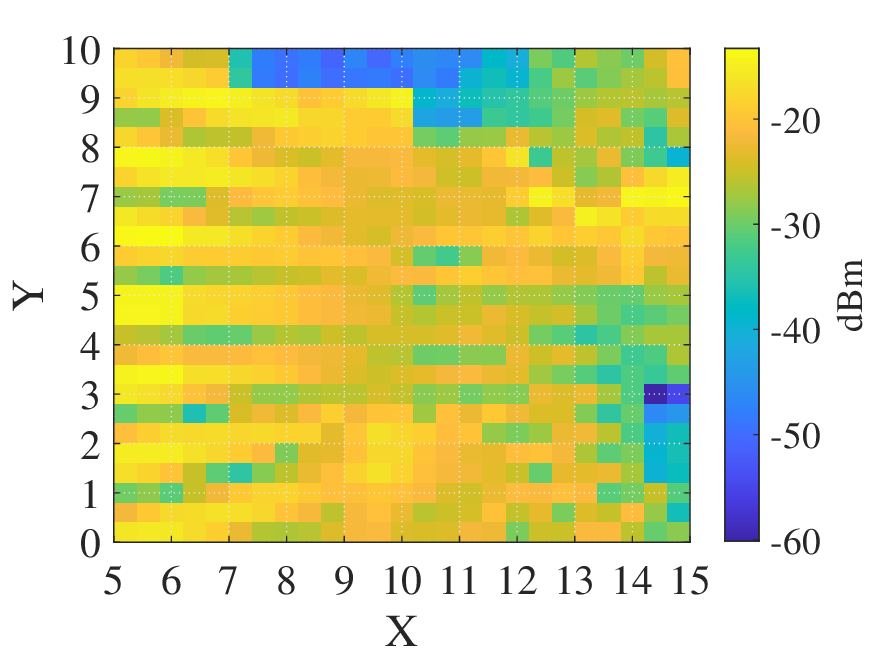}
     }
      \subfloat[ \label{rss4}At the 19-th measurement.]{%
       \includegraphics[width=0.24\textwidth]{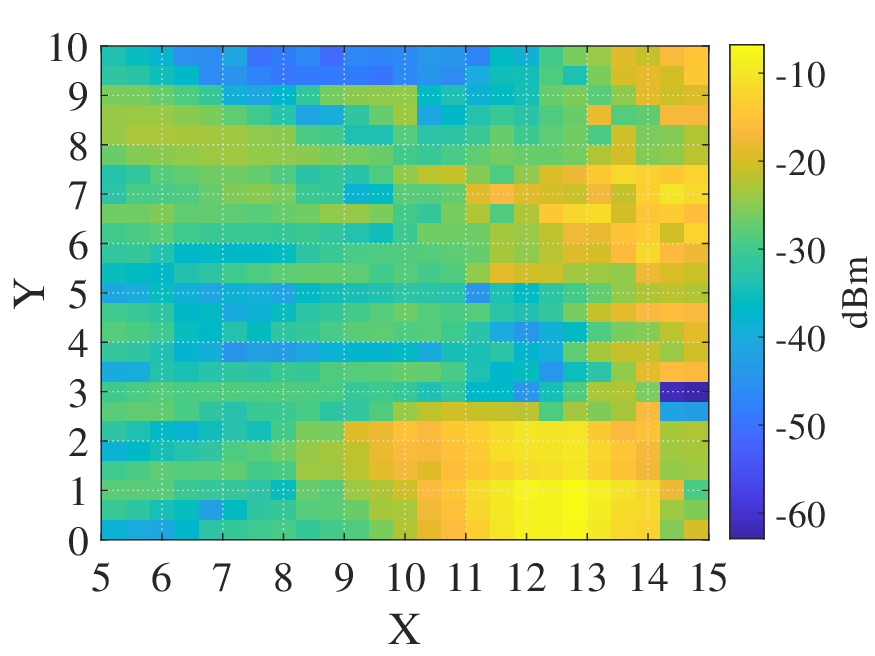}
     }
     \caption{Radio map of RSS values at different measurements.}\label{radiomapfig}
\end{figure*}

Fig.~\ref{radiomapfig} shows the radio map over the AoI at different measurements in RIS-WFL-DB. According to the applied UFSS, the phase shifts of RIS change over the measurements, thus resulting in different RSS distributions over the AoI, as observed in Fig.~\ref{radiomapfig}. Moreover, under the RIS configuration at the a single measurement, the RSS values at distant positions may in close proximity. However, when more RIS configurations are taken, the RSS value sequences at distant positions are unlikely to be close. Therefore, with more RIS configurations, i.e., measurements,  the spatial resolution of wireless fingerprint can be improved.

\subsection{Localization Performance}

\begin{figure}
  \centering
  \includegraphics[width=0.5\textwidth]{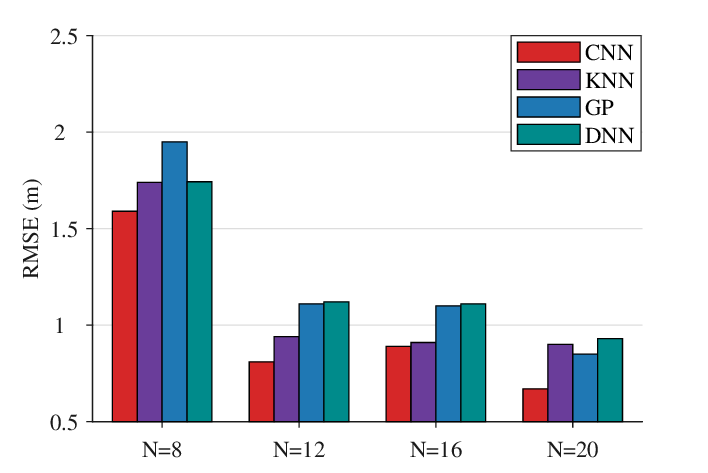}\\
  \caption{RMSEs of different localization algorithms versus number of measurements.}\label{measurmentmethodfig}
\end{figure}

Fig. \ref{measurmentmethodfig} depicts the mean square errors (RMSE) performance under four common fingerprint localization algorithms. It can be observed that, given a sufficient number of measurements, the RMSE of all four algorithms can be reduced to below 1 m. Among them, CNN consistently achieves the lowest error. Particularly when $N=20$, the RMSE of CNN is $0.67$ m, whereas that of the next best performer, GP, is $0.85$ m. This represents a reduction of $21.18\%$ in RMSE for CNN, attributable to its superior capability in feature extraction and noise suppression compared to other methods. Furthermore, according to this figure, the average RMSEs of CNN and KNN are more favorable compared to the other algorithms. To ensure a clear and focused comparative analysis in subsequent simulations, the CNN and KNN are selected as the localization algorithms to present RMSEs.

\begin{figure}
  \centering
  \includegraphics[width=0.5\textwidth]{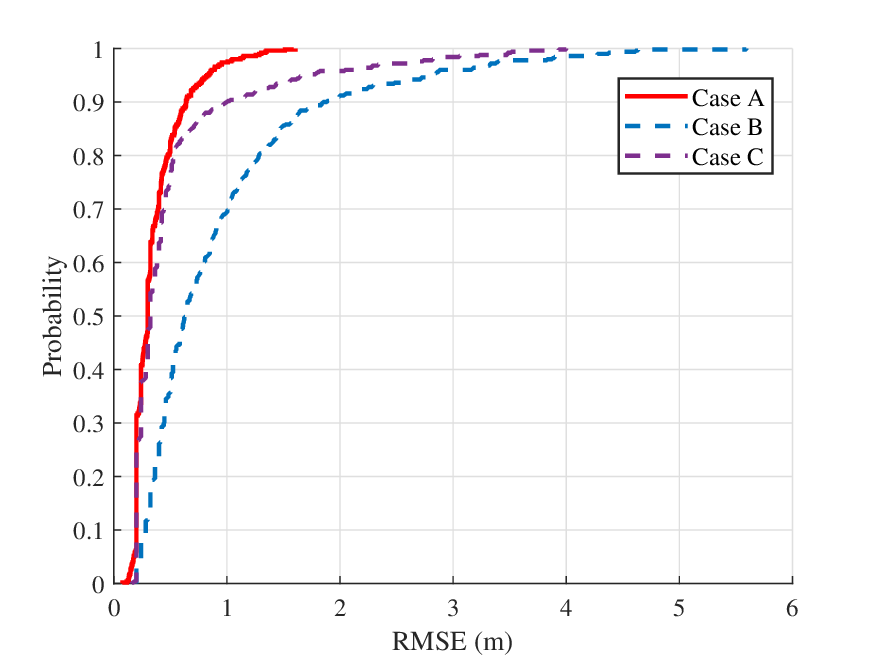}\\
  \caption{CDF of location error under different database generation cases.}\label{casecomparefig}
\end{figure}

Fig.~\ref{casecomparefig} investigates cumulative distribution function (CDF) of location error  across three different database generation cases, named case A, case B and case C. In case A, RIS-WFL-DB is used. In case B, a fingerprint database is generated without RIS, and the other channel parameters are the same as those in RIS-WFL-DB. In case C, a fingerprint database is generated without the spatial consistency procedure of LoS/VLoS conditions and SF terms. And the other channel parameters are the same as those in RIS-WFL-DB. The location error refers to the smaller values from  CNN and KNN algorithms. Comparing case A and case B in Fig.~\ref{casecomparefig}, it is found that incorporating a RIS to indoor WFL can reduce the location error. Moreover, it is found that the location error in case A is much lower than that in case C. Because with spatial consistency procedure, the RSS variation in the AoI becomes smooth, and this spatial consistency also provides more useful information to CNN or KNN. It indicates that when spatial consistency is not used, the positioning accuracy evaluation of  RIS-assisted WFL obviously diverges from the actual scenario.

\begin{figure}
   \subfloat[ \label{rmfig}Reflection magnitude.]{%
       \includegraphics[width=0.25\textwidth]{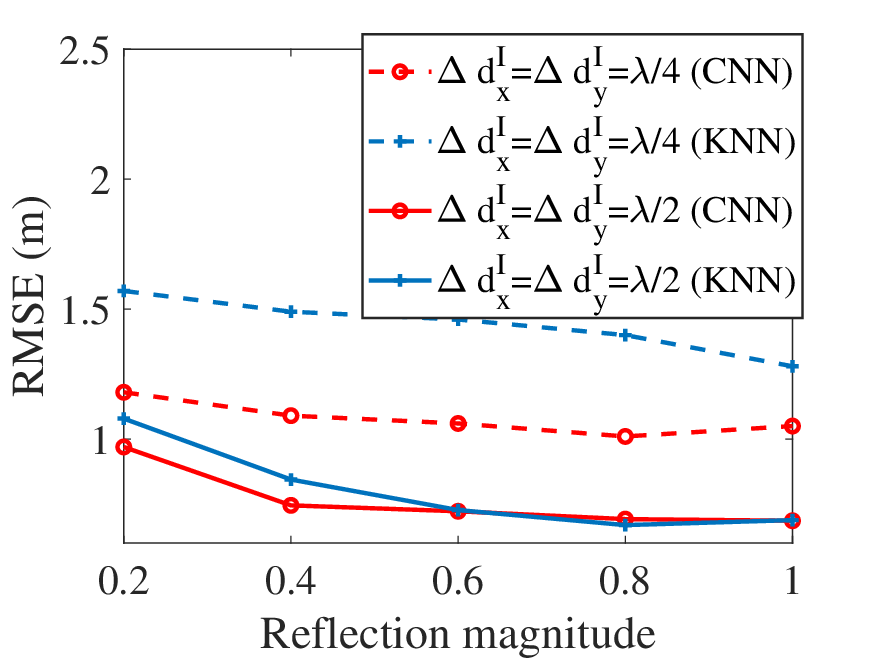}
     }
      \subfloat[ \label{nrufig}Number of reflective units.]{%
       \includegraphics[width=0.25\textwidth]{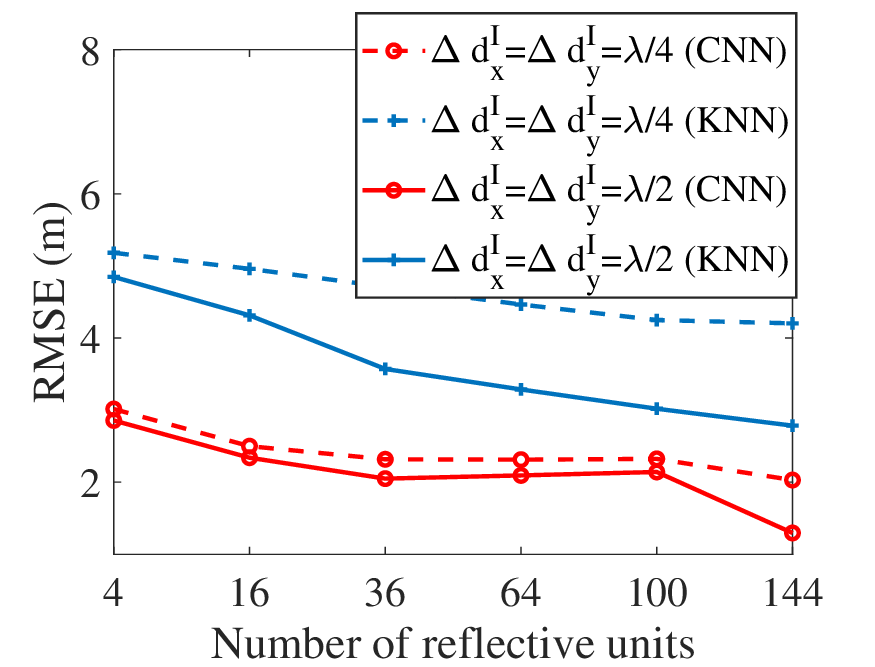}
     }
     \caption{RMSE versus physical factors of RIS.}\label{physicalrisfig}
\end{figure}

Fig.~\ref{physicalrisfig}(a) shows the RMSE versus reflection magnitude of reflective unit under different unit sizes. As expected, both the CNN and KNN exhibit a lower RMSE with larger reflection magnitudes, as the power of VLoS paths increases, which dominates the diversity of the radio maps. This trend is particularly evident in the region with low reflection magnitudes. Additionally, Fig.~\ref{physicalrisfig}(b) illustrates the RMSE versus number of reflective units under different unit sizes. The results indicate that as the number of reflective units increases, the RMSEs of both CNN and KNN decrease, owing to the increased diversity of radio maps. Moreover, from both figures, we can conclude that a larger unit size leads to a lower RMSE.

\begin{figure}
   \subfloat[ \label{clusteramountfig}Numbers of clusters.]{%
       \includegraphics[width=0.25\textwidth]{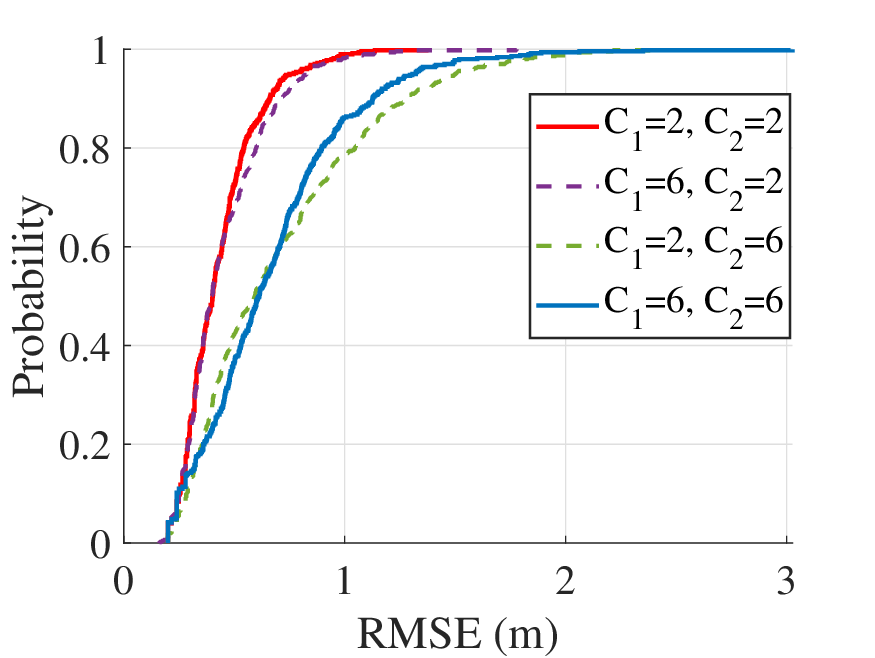}
     }
      \subfloat[ \label{clusterdistancefig}Coherence distances of clusters.]{%
       \includegraphics[width=0.25\textwidth]{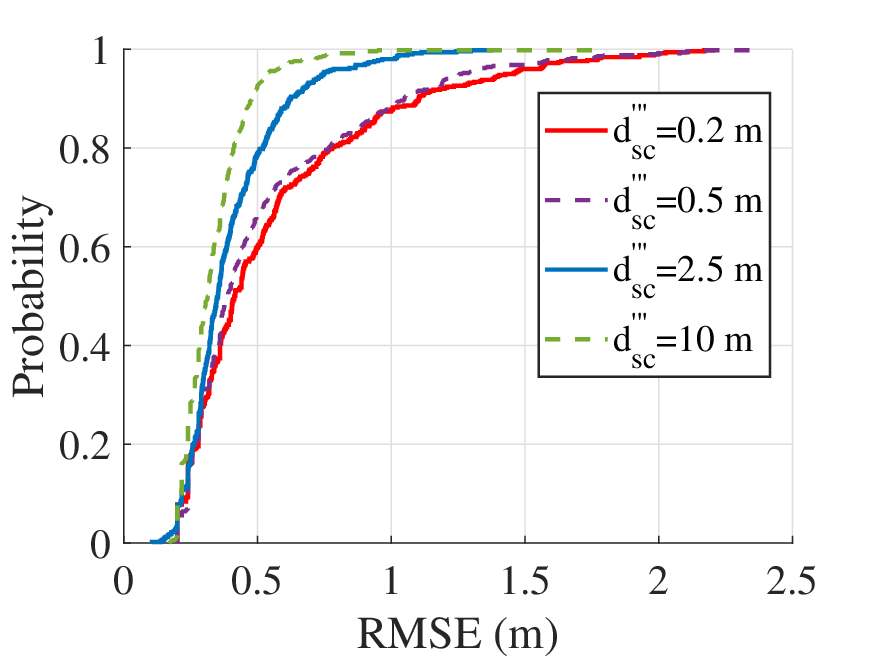}
     }
     \caption{CDF of location error under different numbers and coherence distances of clusters.}\label{clusterfig}
\end{figure}

Fig.~\ref{clusterfig}(a) illustrates the impact of the number of clusters on the distribution of location errors. In this figure, the location error refers to the smaller values from CNN and KNN algorithms. $C_{1}$ represents the average number of clusters on the SB-NLoS paths within the AoI, while $C_{2}$ denotes the average number of clusters on the DB-NLoS paths within the AoI. By comparing the results pairwise, it can be observed that the number of clusters on SB-NLoS paths has a little effect on the location error, whereas the number of clusters on DB-NLoS paths has a large impact. Moreover, the more clusters present on DB-NLoS paths, the larger the location error becomes. For example, when $C_{1}=2$ and $C_{2}=2$, the location error at the $90\%$  is $0.66$ m, when $C_{1}=2$ and $C_{2}=6$, the location error at the $90\%$  increases to 1.24 m. This phenomenon arises from the diversity of radio maps generated by different RIS modulations, making clusters on DB-NLoS paths more influential on the diversity than SB-NLoS paths.

Fig.~\ref{clusterfig}(b) illustrates the impact of coherence distance of clusters on location error, where the location error refers to the smaller values from  CNN and KNN algorithms. As expected, the larger the coherence distance, the smaller the location error. For example, when $d^{'''}_{sc} = 0.5$ m, the location error at the 90\% percentile is 1.03 m while when $d^{'''}_{sc} = 5$ m, the location error at the $90\%$  remains 0.64 m. This is because as the coherence distance increases, the granularity of the spatially correlated map becomes larger, leading to smoother dynamic changes in clusters. This, in turn, allows for better fitting of wireless propagation, thereby reducing location error.

\begin{figure}
  \centering
  \includegraphics[width=0.5\textwidth]{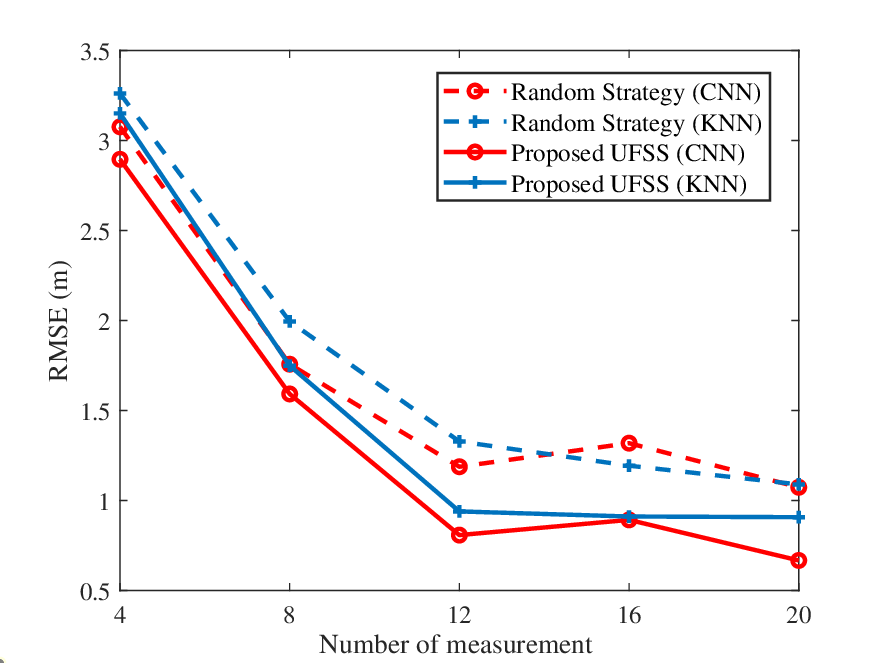}\\
  \caption{RMSEs under different phase configuration strategies versus number of measurements.}\label{rmsestrategyfig}
\end{figure}

Fig.~\ref{rmsestrategyfig} illustrates RMSEs under the proposed UFSS and the basic random strategy versus  number of measurements. The random strategy refers to the phase shifts of RIS being randomly set. It is evident that the proposed UFSS achieves lower RMSE compared to the random strategy. For instance, when $N=12$, the best RMSE of the random strategy is 1.18 m, whereas the best RMSE under the proposed UFSS is 0.81 m. Moreover, the performances of the two strategies are similar only when $N$ is small but becomes discriminative with large $N$. Because a small number of measurements does not provide enough diversity in the radio maps generated by UFSS.  Moreover, as expected, the RMSEs under both strategies decrease with a larger number of measurements.

\begin{figure}
  \centering
  \includegraphics[width=0.5\textwidth]{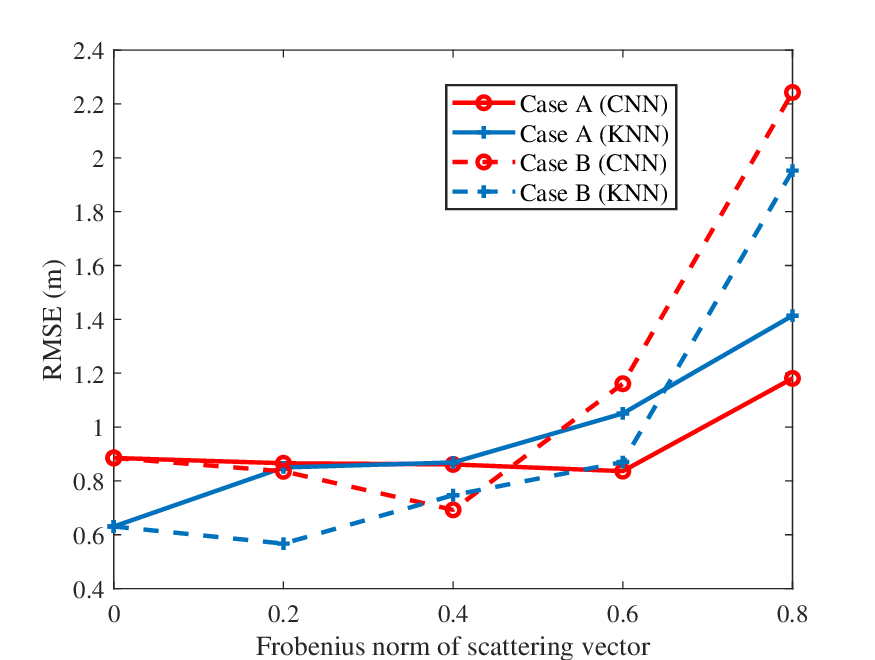}\\
  \caption{RMSE versus Frobenius norm of scattering vector.}\label{couplingfig}
\end{figure}

To investigate the influence of MC on the location error of RIS-assisted WFL, two cases named case A and case B are considered. In case A, the scattering vector collecting scattering values is $[0.606+0.308j,-0.234+0.614j,0.0375-0.324j]$ \cite{fadakar2025mutual}. In case B, the scattering vector is $[-0.681+0.458j,-0.506+0.0492j,0.244+0.0928j]$ \cite{Fadakar2025NearFieldRL}. Fig.~\ref{couplingfig} plots the RMSE as a function of the Frobenius norm of the scattering vector for both cases. The results show that when the Frobenius norm is small, the MC only marginally affects the RMSE. However, as the Frobenius norm increases, the MC significantly increases the RMSE. This is because the MC alters the designed phase configurations, as described by Eq.~(\ref{mcformula}), thereby reducing the diversity of radio maps.

The  precision of the proposed sorting algorithm versus
training and test data size ratio is shown in Table.~\ref{tabesyn}. To test the algorithm, each RSS value sequence in the test database is random disordered. From the table, it can be seen that the proposed sorting algorithm achieves an accuracy rate of over 95.0\%, which solves the synchronization issue well.

\begin{table}
\centering
\captionsetup{justification=centering, labelsep=newline,textfont=sc}
\caption{Synchronization Precision of Proposed Sorting Algorithm}
\label{tabesyn}
\begin{tabular}{p{0.2\textwidth}|p{0.03\textwidth}|p{0.03\textwidth}|p{0.03\textwidth}|p{0.03\textwidth}|p{0.03\textwidth}}
\hline
\textbf{Data Size Ratio} & 9:1 & 8:2 & 7:3 & 6:4 & 5:5 \\
\hline
\textbf{Synchronization Precision}  & 98.8\% & 97.4\% & 96.4\% &95.7\% & 95.0\% \\
\hline
\end{tabular}
\end{table}

\subsection{Effectiveness Verification}
\begin{figure}
   \subfloat[ \label{floorplanfig}A 2-D view of an indoor office.]{%
       \includegraphics[width=0.25\textwidth]{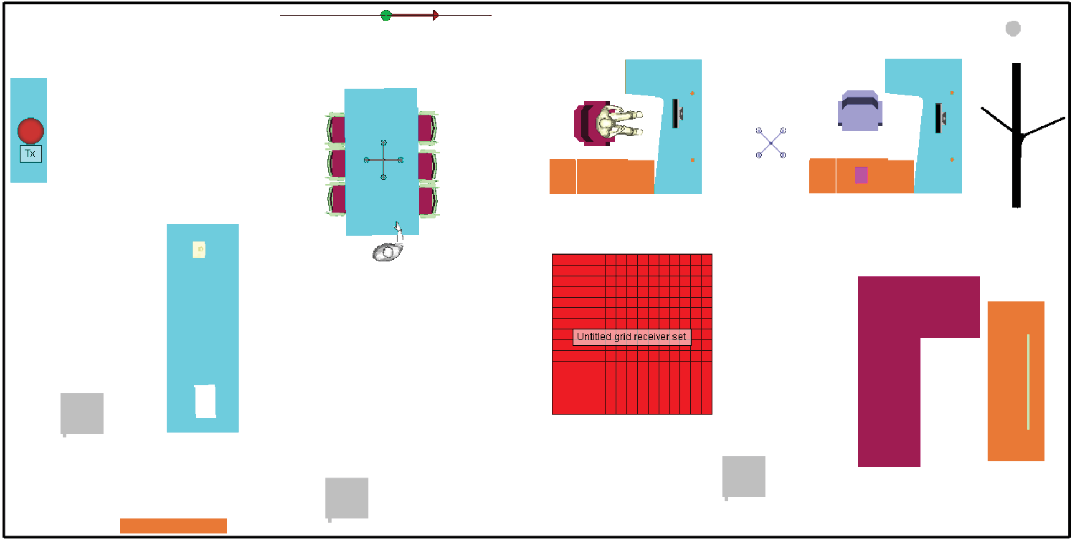}
     }
      \subfloat[ \label{clusterdistancefig}Receive powers of ten Rx positions.]{%
       \includegraphics[width=0.25\textwidth]{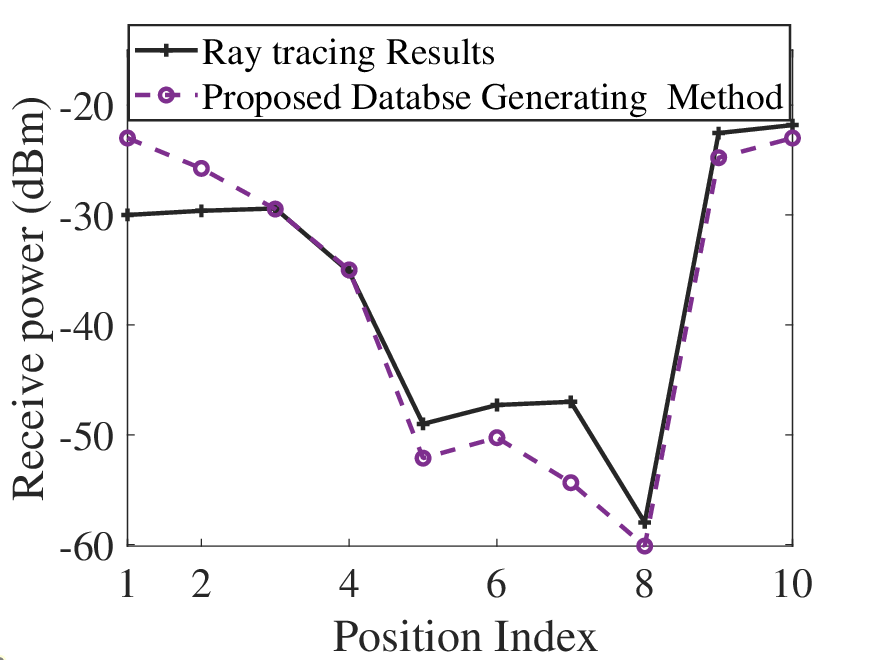}
     }
     \caption{Verification of the proposed database generation method.}\label{wirelessinsiteverfiyfig}
\end{figure}

To verify the effectiveness of the proposed database generation method, the ray tracing method using Wireless InSite software is used as the benchmark, which is widely used in wireless propagation studies \cite{liu2025indoor}. Fig.~\ref{wirelessinsiteverfiyfig}(a) shows a 2-D view of the established indoor office wireless propagation environment in Wireless InSite. The red circle on the left table is the projection of Tx, the red line with arrow is the projection of RIS, and the red grids are the projection of Rx positions. The ray tracing results in the Wireless InSite are compared with the proposed database generation method where the channel parameters are estimated using the DNN approach \cite{liu2020survey}.

Fig.~\ref{wirelessinsiteverfiyfig}(b) plots the receive powers of ten Rx positions under the ray tracing method and the proposed database generation method. These points are spaced 0.2 m apart. In this figure, positions with indexes 5, 6, and 7 are in NLoS scenarios while the other positions are in LoS/VLoS scenarios. Therefore lower receive powers are observed in positions with indexes 5, 6, and 7. This figure also indicates the spatial consistency of wireless propagation such as the continuity of the LoS/VLoS condition and the gradual variation of receive power in specific LoS/VLoS and NLoS environments. Moreover, the deviation between the proposed database generation method and the ray tracing method is within 3 dBm in most positions,  which demonstrate the realism of the proposed method.

\section{Conclusion}
In this paper, we propose a novel fingerprint database generation method for RIS-assisted WFL. First, we introduce a general and flexible cluster-based channel modeling approach to simulate RIS-assisted wireless propagation. Next, a spatial consistency procedure is proposed to capture spatially consistent channel parameters. Finally, the process for generating the fingerprint database is outlined. Furthermore, important practical application issues are resolved including online synchronization, RIS phase configuration and MC effect. Extensive simulations are performed on the generated fingerprint database, and the results demonstrate the significant potential of incorporating RIS to enhance WFL. The proposed database generation method will facilitate future research on RIS-assisted WFL.

\appendices

\ifCLASSOPTIONcaptionsoff
  \newpage
\fi
\small
\bibliography{IEEEfull,cite}

\begin{thebibliography}{10}
\providecommand{\url}[1]{#1}
\csname url@samestyle\endcsname
\providecommand{\newblock}{\relax}
\providecommand{\bibinfo}[2]{#2}
\providecommand{\BIBentrySTDinterwordspacing}{\spaceskip=0pt\relax}
\providecommand{\BIBentryALTinterwordstretchfactor}{4}
\providecommand{\BIBentryALTinterwordspacing}{\spaceskip=\fontdimen2\font plus
\BIBentryALTinterwordstretchfactor\fontdimen3\font minus
  \fontdimen4\font\relax}
\providecommand{\BIBforeignlanguage}[2]{{%
\expandafter\ifx\csname l@#1\endcsname\relax
\typeout{** WARNING: IEEEtran.bst: No hyphenation pattern has been}%
\typeout{** loaded for the language `#1'. Using the pattern for}%
\typeout{** the default language instead.}%
\else
\language=\csname l@#1\endcsname
\fi
#2}}
\providecommand{\BIBdecl}{\relax}
\BIBdecl

\bibitem{zafari2019survey}
F.~Zafari, A.~Gkelias, and K.~K. Leung, ``A survey of indoor localization
  systems and technologies,'' \emph{IEEE Communications Surveys \& Tutorials},
  vol.~21, no.~3, pp. 2568--2599, 2019.

\bibitem{nkrow2024nlos}
R.~E. Nkrow, B.~Silva, D.~Boshoff, G.~Hancke, M.~Gidlund, and A.~Abu-Mahfouz,
  ``{NLOS} identification and mitigation for time-based indoor localization
  systems: survey and future research directions,'' \emph{ACM Computing
  Surveys}, vol.~56, no.~12, pp. 1--41, 2024.

\bibitem{singh2024machine}
A.~Singh, V.~Kalaichelvi, and R.~Karthikeyan, ``Machine learning-based
  multi-sensor fusion for warehouse robot in {GPS}-denied environment,''
  \emph{Multimedia Tools and Applications}, vol.~83, no.~18, pp.
  56\,229--56\,246, 2024.

\bibitem{shastri2022review}
A.~Shastri, N.~Valecha, E.~Bashirov, H.~Tataria, M.~Lentmaier, F.~Tufvesson,
  M.~Rossi, and P.~Casari, ``A review of millimeter wave device-based
  localization and device-free sensing technologies and applications,''
  \emph{IEEE Communications Surveys \& Tutorials}, vol.~24, no.~3, pp.
  1708--1749, 2022.

\bibitem{gonzalez2024integrated}
N.~Gonz{\'a}lez-Prelcic, M.~F. Keskin, O.~Kaltiokallio, M.~Valkama, D.~Dardari,
  X.~Shen, Y.~Shen, M.~Bayraktar, and H.~Wymeersch, ``The integrated sensing
  and communication revolution for {6G}: Vision, techniques, and
  applications,'' \emph{Proceedings of the IEEE}, vol. 112, no.~7, pp.
  676--723, 2024.

\bibitem{cheng2022federated}
X.~Cheng, C.~Ma, J.~Li, H.~Song, F.~Shu, and J.~Wang, ``Federated
  learning-based localization with heterogeneous fingerprint database,''
  \emph{IEEE Wireless Communications Letters}, vol.~11, no.~7, pp. 1364--1368,
  2022.

\bibitem{e2025systematic}
E.~J.~C. e~Silva, K.~Y. G.~V. Guedes, P.~R. de~Almeida~Ribeiro, A.~O.
  Barradas~Filho \emph{et~al.}, ``Systematic review of radio wave techniques
  for indoor positioning systems,'' \emph{IEEE Latin America Transactions},
  vol.~23, no.~3, pp. 205--215, 2025.

\bibitem{wu2023fingerprint}
T.~Wu, C.~Pan, Y.~Pan, H.~Ren, M.~Elkashlan, and C.-X. Wang,
  ``Fingerprint-based mmwave positioning system aided by reconfigurable
  intelligent surface,'' \emph{IEEE Wireless Communications Letters}, vol.~12,
  no.~8, pp. 1379--1383, 2023.

\bibitem{wu2019towards}
Q.~Wu and R.~Zhang, ``Towards smart and reconfigurable environment: Intelligent
  reflecting surface aided wireless network,'' \emph{IEEE communications
  magazine}, vol.~58, no.~1, pp. 106--112, 2019.

\bibitem{cheng2021joint}
X.~Cheng, Y.~Lin, W.~Shi, J.~Li, C.~Pan, F.~Shu, Y.~Wu, and J.~Wang, ``Joint
  optimization for {RIS}-assisted wireless communications: From physical and
  electromagnetic perspectives,'' \emph{IEEE Transactions on Communications},
  vol.~70, no.~1, pp. 606--620, 2021.

\bibitem{shu2024precoding}
F.~Shu, Y.~Wang, X.~Wang, G.~Xia, L.~Yang, W.~Shi, C.~Shen, and J.~Wang,
  ``Precoding and beamforming design for intelligent reconfigurable
  surface-aided hybrid secure spatial modulation,'' \emph{IEEE Transactions on
  Wireless Communications}, vol.~23, no.~9, pp. 11\,770--11\,784, 2024.

\bibitem{emenonye2023fundamentals}
D.-R. Emenonye, H.~S. Dhillon, and R.~M. Buehrer, ``Fundamentals of {RIS}-aided
  localization in the far-field,'' \emph{IEEE Transactions on Wireless
  Communications}, vol.~23, no.~4, pp. 3408--3424, 2023.

\bibitem{wang2021joint}
W.~Wang and W.~Zhang, ``Joint beam training and positioning for intelligent
  reflecting surfaces assisted millimeter wave communications,'' \emph{IEEE
  Transactions on Wireless Communications}, vol.~20, no.~10, pp. 6282--6297,
  2021.

\bibitem{pan2023ris}
Y.~Pan, C.~Pan, S.~Jin, and J.~Wang, ``{RIS}-aided near-field localization and
  channel estimation for the terahertz system,'' \emph{IEEE Journal of Selected
  Topics in Signal Processing}, vol.~17, no.~4, pp. 878--892, 2023.

\bibitem{zhang2023approximate}
W.~Zhang, Z.~Wang, and W.~P. Tay, ``Approximate maximum-likelihood {RIS}-aided
  positioning,'' \emph{IEEE Transactions on Wireless Communications}, vol.~22,
  no.~12, pp. 8859--8875, 2023.

\bibitem{nguyen2021wireless}
C.~L. Nguyen, O.~Georgiou, G.~Gradoni, and M.~Di~Renzo, ``Wireless
  fingerprinting localization in smart environments using reconfigurable
  intelligent surfaces,'' \emph{IEEE access}, vol.~9, pp. 135\,526--135\,541,
  2021.

\bibitem{zhang2022multiple}
Z.~Zhang, L.~Wu, J.~Dang, B.~Zhu, and L.~Wang, ``Multiple {RSS} fingerprint
  based indoor localization in {RIS}-assisted {5G} wireless communication
  system,'' \emph{The International Archives of the Photogrammetry, Remote
  Sensing and Spatial Information Sciences}, vol.~46, pp. 287--292, 2022.

\bibitem{wang2023intelligent}
Y.~Wang, I.~W.-H. Ho, S.~Zhang, and Y.~Wang, ``Intelligent reflecting surface
  enabled fingerprinting-based localization with deep reinforcement learning,''
  \emph{IEEE Transactions on Vehicular Technology}, vol.~72, no.~10, pp.
  13\,162--13\,172, 2023.

\bibitem{sardellitti2024ris}
S.~Sardellitti, P.~Di~Lorenzo, and S.~Barbarossa, ``{RIS}-aided wireless
  fingerprinting localization based on multilayer graph representations,''
  \emph{IEEE Communications Letters}, vol.~28, no.~5, pp. 1043--1047, 2024.

\bibitem{javed2024fingerprinting}
A.~Javed, N.~U. Hassan, A.~Rafique, M.~Zubair, M.~Di~Renzo, and C.~Yuen,
  ``Fingerprinting database development methods for reconfigurable intelligent
  surface assisted indoor positioning system,'' \emph{IEEE Access}, vol.~12,
  pp. 85\,244--85\,258, 2024.

\bibitem{hou2025indoor}
Y.~Hou, K.~Katsuma, S.~Tada, T.~Kaito, and S.~Denno, ``Indoorlocalization for
  mmwave wireless system using diversity property of {RIS} device,'' in
  \emph{2025 IEEE 8th Information Technology and Mechatronics Engineering
  Conference (ITOEC)}, vol.~8.\hskip 1em plus 0.5em minus 0.4em\relax IEEE,
  2025, pp. 1412--1416.

\bibitem{3GPP_TR_38_901}
\BIBentryALTinterwordspacing
{3GPP}, ``Study on channel model for frequencies from 0.5 to 100 {GHz},
  v18.0.0,'' Sophia Antipolis, France, Technical Report, Mar. 2024. [Online].
  Available:
  \url{https://portal.3gpp.org/desktopmodules/Specifications/SpecificationDetails.aspx?specificationId=3173}
\BIBentrySTDinterwordspacing

\bibitem{poddar2023tutorial}
H.~Poddar, S.~Ju, D.~Shakya, and T.~S. Rappaport, ``A tutorial on {NYUSIM}:
  Sub-terahertz and millimeter-wave channel simulator for {5G}, {6G}, and
  beyond,'' \emph{IEEE Communications Surveys \& Tutorials}, vol.~26, no.~2,
  pp. 824--857, 2023.

\bibitem{basar2021indoor}
E.~Basar, I.~Yildirim, and F.~Kilinc, ``Indoor and outdoor physical channel
  modeling and efficient positioning for reconfigurable intelligent surfaces in
  mmwave bands,'' \emph{IEEE Transactions on Communications}, vol.~69, no.~12,
  pp. 8600--8611, 2021.

\bibitem{xiong2021statistical}
B.~Xiong, Z.~Zhang, H.~Jiang, H.~Zhang, J.~Zhang, L.~Wu, and J.~Dang, ``A
  statistical {MIMO} channel model for reconfigurable intelligent surface
  assisted wireless communications,'' \emph{IEEE Transactions on
  Communications}, vol.~70, no.~2, pp. 1360--1375, 2021.

\bibitem{lian2024physics}
Z.~Lian, W.~Zhang, Y.~Wang, Y.~Su, B.~Zhang, B.~Jin, and B.~Wang,
  ``Physics-based channel modeling for {IRS}-assisted mmwave communication
  systems,'' \emph{IEEE Transactions on Communications}, vol.~72, no.~5, pp.
  2687--2700, 2024.

\bibitem{yuan2024ris}
Y.~Yuan, R.~He, B.~Ai, Z.~Zhang, and Y.~Jin, ``{RIS}-assisted mobile channels
  with directional transmission: Modeling and characteristic analysis,''
  \emph{IEEE Transactions on Wireless Communications}, vol.~23, no.~11, pp.
  17\,021--17\,036, 2024.

\bibitem{alitaleshi2023ea}
A.~Alitaleshi, H.~Jazayeriy, and J.~Kazemitabar, ``{EA-CNN}: A smart indoor
  {3D} positioning scheme based on {Wi-Fi} fingerprinting and deep learning,''
  \emph{Engineering Applications of Artificial Intelligence}, vol. 117, p.
  105509, 2023.

\bibitem{zeng2020reconfigurable}
S.~Zeng, H.~Zhang, B.~Di, Z.~Han, and L.~Song, ``Reconfigurable intelligent
  surface ({RIS}) assisted wireless coverage extension: {RIS} orientation and
  location optimization,'' \emph{IEEE Communications Letters}, vol.~25, no.~1,
  pp. 269--273, 2020.

\bibitem{hemadeh2017millimeter}
I.~A. Hemadeh, K.~Satyanarayana, M.~El-Hajjar, and L.~Hanzo, ``Millimeter-wave
  communications: Physical channel models, design considerations, antenna
  constructions, and link-budget,'' \emph{IEEE Communications Surveys \&
  Tutorials}, vol.~20, no.~2, pp. 870--913, 2017.

\bibitem{molisch2012wireless}
A.~F. Molisch, \emph{Wireless communications}.\hskip 1em plus 0.5em minus
  0.4em\relax John Wiley \& Sons, 2012.

\bibitem{tang2022path}
W.~Tang, X.~Chen, M.~Z. Chen, J.~Y. Dai, Y.~Han, M.~Di~Renzo, S.~Jin, Q.~Cheng,
  and T.~J. Cui, ``Path loss modeling and measurements for reconfigurable
  intelligent surfaces in the millimeter-wave frequency band,'' \emph{IEEE
  Transactions on Communications}, vol.~70, no.~9, pp. 6259--6276, 2022.

\bibitem{docomo20165g}
\BIBentryALTinterwordspacing
N.~Docomo \emph{et~al.}, ``{5G} channel model for bands up to 100 {GHz},''
  2016. [Online]. Available:
  \url{http://www.5gworkshops.com/5GCMSIG_White%20Paper_r2dot3.pdf}
\BIBentrySTDinterwordspacing

\bibitem{jiang20213gpp}
T.~Jiang, J.~Zhang, P.~Tang, L.~Tian, Y.~Zheng, J.~Dou, H.~Asplund,
  L.~Raschkowski, R.~D’Errico, and T.~J{\"a}ms{\"a}, ``{3GPP} standardized
  {5G} channel model for {IIoT} scenarios: A survey,'' \emph{IEEE Internet of
  Things Journal}, vol.~8, no.~11, pp. 8799--8815, 2021.

\bibitem{Cui2025RIS}
\BIBentryALTinterwordspacing
T.~J. Cui, J.~H. Zhang, J.~W. Dou, S.~Jin, Y.~F. Yuan, Q.~Y. Liu, N.~X. Li, and
  {et al.}, ``Channel modeling and simulation for reconfigurable intelligent
  surface,'' FuTURE Forum, Nanjing, China, Technical Report, Apr. 2025.
  [Online]. Available: \url{https://doi.org/10.12142/FuTURE.202504004}
\BIBentrySTDinterwordspacing

\bibitem{sang2023multi}
J.~Sang, M.~Zhou, J.~Lan, B.~Gao, W.~Tang, X.~Li, S.~Jin, E.~Basar, C.~Li,
  Q.~Cheng \emph{et~al.}, ``Multi-scenario broadband channel measurement and
  modeling for sub-6 {GHz} {RIS}-assisted wireless communication systems,''
  \emph{IEEE Transactions on Wireless Communications}, vol.~23, no.~6, pp.
  6312--6329, 2023.

\bibitem{balanis2016antenna}
C.~A. Balanis, \emph{Antenna theory: analysis and design}.\hskip 1em plus 0.5em
  minus 0.4em\relax John wiley \& sons, 2016.

\bibitem{jamali2022low}
V.~Jamali, G.~C. Alexandropoulos, R.~Schober, and H.~V. Poor,
  ``Low-to-zero-overhead {IRS} reconfiguration: Decoupling illumination and
  channel estimation,'' \emph{IEEE Communications Letters}, vol.~26, no.~4, pp.
  932--936, 2022.

\bibitem{xu2024swin}
X.~Xu, F.~Zhu, S.~Han, Z.~Yu, H.~Zhao, B.~Wang, and P.~Zhang, ``Swin-loc:
  Transformer-based {CSI} fingerprinting indoor localization with {MIMO} {ISAC}
  system,'' \emph{IEEE Transactions on Vehicular Technology}, vol.~73, no.~8,
  pp. 11\,664--11\,679, 2024.

\bibitem{fadakar2025mutual}
A.~Fadakar, M.~F. Keskin, H.~Chen, and H.~Wymeersch, ``Mutual coupling-aware
  localization for {RIS}-assisted {ISAC} systems,'' \emph{IEEE Transactions on
  Cognitive Communications and Networking}, vol.~11, no.~5, pp. 2938--2954,
  2025.

\bibitem{zheng2024mutual}
P.~Zheng, R.~Wang, A.~Shamim, and T.~Y. Al-Naffouri, ``Mutual coupling in
  {RIS}-aided communication: Model training and experimental validation,''
  \emph{IEEE Transactions on Wireless Communications}, 2024.

\bibitem{Fadakar2025NearFieldRL}
A.~Fadakar, M.~F. Keskin, H.~Chen, H.~Wymeersch, and A.~F. Molisch,
  ``Near-field {RIS}-assisted localization under mutual coupling,'' in
  \emph{2025 IEEE International Conference on Communications Workshops (ICC
  Workshops)}.\hskip 1em plus 0.5em minus 0.4em\relax IEEE, 2025, pp.
  1470--1475.

\bibitem{ibrahim2018cnn}
M.~Ibrahim, M.~Torki, and M.~ElNainay, ``{CNN} based indoor localization using
  {RSS} time-series,'' in \emph{2018 IEEE symposium on computers and
  communications (ISCC)}.\hskip 1em plus 0.5em minus 0.4em\relax IEEE, 2018,
  pp. 01\,044--01\,049.

\bibitem{kumar2016gaussian}
S.~Kumar, R.~M. Hegde, and N.~Trigoni, ``Gaussian process regression for
  fingerprinting based localization,'' \emph{Ad Hoc Networks}, vol.~51, pp.
  1--10, 2016.

\bibitem{liu2025indoor}
L.~Liu, Z.~Tian, Y.~Wang, Y.~Jin, L.~Li, and S.~Gui, ``Indoor environment
  mapping and localization based on a single {Wi-Fi} access point,'' in
  \emph{ICC 2025-IEEE International Conference on Communications}.\hskip 1em
  plus 0.5em minus 0.4em\relax IEEE, 2025, pp. 2503--2508.

\bibitem{liu2020survey}
L.~Liu, J.~Zhang, Y.~Fan, Y.~Li, and J.~Zhang, ``Survey of application of
  machine learning in wireless channel modeling,'' \emph{Journal on
  Communications}, vol.~42, no.~3, pp. 134--153, 2020.

\end{thebibliography}
\bibliographystyle{IEEEtran}
\end{document}